\RequirePackage[hyphens]{url}
\PassOptionsToPackage{round,authoryear}{natbib}
\documentclass[reprint,
nofootinbib,
 amsmath,amssymb,
 aps,
pra
]{revtex4-1}
\usepackage{url,hyperref,lineno,microtype,subcaption,listings,xcolor,graphicx}
\usepackage{booktabs}

\definecolor{codegreen}{rgb}{0,0.6,0}
\definecolor{codegray}{rgb}{0.5,0.5,0.5}
\definecolor{codepurple}{rgb}{0.58,0,0.82}
\definecolor{backcolour}{rgb}{0.95,0.95,0.95}

\lstdefinestyle{mystyle}{
    backgroundcolor=\color{backcolour},   
    commentstyle=\color{codegreen},
    keywordstyle=\color{magenta},
    numberstyle=\tiny\color{codegray},
    stringstyle=\color{codepurple},
    basicstyle=\ttfamily,
    breakatwhitespace=false,         
    breaklines=true,                 
    captionpos=b,                    
    keepspaces=true,                 
    numbers=left,                    
    numbersep=5pt,                  
    showspaces=false,                
    showstringspaces=false,
    showtabs=false,                  
    tabsize=2
}

\usepackage{orcidlink}

\bibpunct{(}{)}{;}{a}{,}{,}

\begin{document}
\title{The Rise and Fall of the Initial Era}

\author{Simon J Porter \orcidlink{0000-0002-6151-8423} 0000-0002-6151-8423}
 \email{s.porter@digital-science.com}
  
 \affiliation{%
 Digital Science, 6 Briset Street, London, EC1M 5NR, UK
}
\author{Daniel W Hook \orcidlink{0000-0001-9746-1193} 0000-0001-9746-1193}
 \email{daniel@digital-science.com}
  
 \affiliation{%
 Digital Science, 6 Briset Street, London, EC1M 5NR, UK
}

\begin{abstract}
Bibliographic data is a rich source of information that goes beyond the use cases of location and citation---it also encodes both cultural and technological context.  This paper uses large-scale analysis of author-name representation in the \textit{Dimensions} database to identify and characterise an ``Initial Era'' in the scholarly record, running from 1945 to 1983, during which initials were used in preference to full names on scholarly communications. We document this era's emergence, its persistence across countries and disciplines, and its rapid decline from approximately 2002 in conjunction with specific technological and policy changes in the bibliographic infrastructure.  We argue that the Initial Era is exceptional in the four-century history of formalised scholarly communication, and we examine its implications for the visibility of researchers---particularly women---in the scholarly record, and for the broader project of bibliometric archaeology over digital research data.
\end{abstract}
\maketitle

\section{Introduction}
In the contemporary discourse on technology, dominated by references to digital and electronic innovations, the notion of the research article as a form of technology may appear incongruous. However, an exploration of the research article through the lens of technology is critical for a comprehensive understanding of its role, interactions with other technological forms, and its consequent impact on society. The research article, viewed technologically, is a significant construct, with a long-standing history of shaping social norms and establishing institutions that extend their influence across the research community, irrespective of disciplinary boundaries, geographical locations, or historical periods \citep{hall_varieties_2001}. This technology has achieved ubiquity on three distinct levels: spatial, with researchers globally engaging with and understanding research articles under shared assumptions; temporal, allowing for the contextual comprehension of older articles through a slow evolution of the format; and disciplinary, with a cross-disciplinary recognition of the rigorous scrutiny and scientific methodology underpinning the work.  These characteristics are critical for research to function, as an incremental activity that builds on prior results and knowledge.

The intrinsic characteristics of research papers have rendered them the foundational elements of research communication and, crucially, the conduits of trust among researchers, transcending spatial, temporal, and disciplinary divides. This established trust facilitates incremental research and underpins the development and cohesion of the global research community. Analogous to economic institutions, the norms surrounding research papers enable researchers to make assumptions similar to the reliability of contracts in legally robust countries, thus enabling international academic transactions. Beyond facilitating trust, the structured format of a research paper—detailing the research's specifics, the researchers, the location and timing of the research, funding sources, and relevant previous work—supports the provenance and contextualisation essential for the credibility of its communicated results.

The interaction between technology and its consequent influence over its users and communities is a well-documented phenomenon; however, possibly due to the long-lived and slow-changing nature of its underlying format, the research paper stands out for its persistence over the centuries. Over approximately the first 300 years of formalised research communication, dating back to the 1660s, the pace of change in research publication formats has been gradual. Only in the last half-century has the rapid transformation of research practices necessitated a quicker evolution of this technology.

At its core, the research paper remains a rare example of a 17$^{\rm th}$-Century technology in current use.  It is a technology that originally emerged from a very different time to serve a very different context. It was the Age of Empire, where research communications were characterised by brief correspondences, often containing hand-drawn diagrams or concise result tables, exchanged among a small, affluent elite. Originating in the club culture of coffee houses in cities like London and Paris, this technology now underpins a vast, global research ecosystem, supporting millions of contributors and a burgeoning diversity of cultures, geographies, and subjects \citep{jarvis_gutenberg_2023,johns_science_2023,taylor_how_2024}.

The study of the representation of author names provides us with an insight into the structure of research culture and the sociology of research.  This has been well-studied with the implications of gender bias in various aspects of the scholarly record being examined in detail \citep{macaluso_is_2016,ni_gendered_2021,sugimoto_factors_2019,kozlowski_avoiding_2022,kozlowski_intersectional_2022}.  It is not the focus of this article to rehash these arguments but rather to add to the data supporting these arguments.   In a period where there are high proportions of researchers using full-form names, we are able potentially to determine more about the gender and background of the participants in the research ecosystem and thus learn about the geographical and discipline-based diversity of the participants in the research ecosystem.  When we enter periods where initial-form is used, data become less rich and we are able to determine less.

This study leverages the slow evolution of the research paper to conduct a form of ``bibliometric archaeology'' \citep{meadows_1985} over the digital research record \citep{hook_scaling_2021,bornmann_growth_2021}, examining the interplay between technological and societal changes as encapsulated in the history of the research paper's format. By focusing on seemingly minor details, such as the presentation of author names, from full given names (``full form'') to initials only (``initial form''), this paper seeks to uncover broader narratives, correlating shifts in norms with technological, geographic, disciplinary, and societal factors.

The paper is organised as follows: In the remainder of the current section we set a historical backdrop that gives an insight into the consistency of the data that we study.  In the methodology section, we give an overview of the data sources and technology used with exemplar code, and we describe the approach that we have taken to classify authors and papers into the different cohorts required for the analyses described above.  In the results section we present several analyses and interpret these.  Finally, we conclude the paper with a brief discussion of the results and suggest future directions of research.

The contribution of this paper is threefold. First, we identify the existence of an ``Initial Era'' in the scholarly record (1945--1983) during which initial-form author names dominated, and we argue that this era is exceptional rather than typical of the historical record. Second, we trace the rise and fall of this era through complementary analyses at the level of papers, authors, countries, disciplines, journals, and bibliographic technology, in order to disentangle the contributions of cultural, geographic, disciplinary, and technological drivers. Third, we use the Initial Era as a case study in \textit{bibliometric archaeology}---the project of reading the digital scholarly record as a layered artefact whose features reflect not only the activity it records but also the policies, technologies, and conventions that produced it. We return to this framing in the Discussion.

\subsection{A brief history of metadata norms}
The prevailing norm within contemporary academic discourse mandates the association of scholarly outputs---be it journal articles, conference proceedings, books, or other forms---with one or more authors. This practice, however, was not always a staple of scholarly communication. The explicit linkage of authors to their scholarly contributions, particularly when introducing novel results or claims, has evolved into a cornerstone of the modern research paradigm. This evolution is driven largely by the critical role of provenance in establishing the reliability of claims, facilitating subsequent research built upon these foundations, and the necessity for researchers to be credited with their findings as a means of securing funding and institutional support.

Despite the ubiquity of this norm, contemporary scholarly practices do witness exceptions, particularly in types of outputs that do not introduce novel findings, such as editorial articles, where the omission of author names remains more common.

In the nascent stages of formal scholarly communication, particularly within the ``club'' culture that characterised the early operations of the Philosophical Transactions of the Royal Society—regarded as the first formal research journal—the practice of attributing author names to scholarly works was not deemed essential. An analysis of the journal’s publications from its initial five years (1665-1669) reveals that only 171 out of 357 articles listed in the \textit{Dimensions} database from Digital Science feature identifiable authors. Furthermore, during this period, not a single contribution to the journal provided the address or institutional affiliation of the author, underscoring a significant departure from contemporary standards of authorship and attribution.

We acknowledge that this framing is Eurocentric. Scholarly traditions of substantial sophistication existed long before, and in parallel with, the seventeenth-century European learned societies---including in mathematics, astronomy, medicine, philosophy, and statecraft across the Islamic world, China, India, and elsewhere. Our focus on the \textit{Philosophical Transactions} and its successors reflects not a claim that scholarship began in seventeenth-century London, but a narrower observation: the journal article as a \textit{technology}---with the conventions of authorship, attribution, and citation that we examine here---emerges from this particular tradition, and it is this technology, in its modern globalised form, that we are studying. The earlier and parallel traditions of scholarship that produced different communication forms (treatises, commentaries, correspondence networks) lie outside the scope of the present analysis but are an important reminder that the conventions we document are historically contingent rather than universal.

\begin{figure}[h!]
\centering
\includegraphics[width=.95\linewidth]{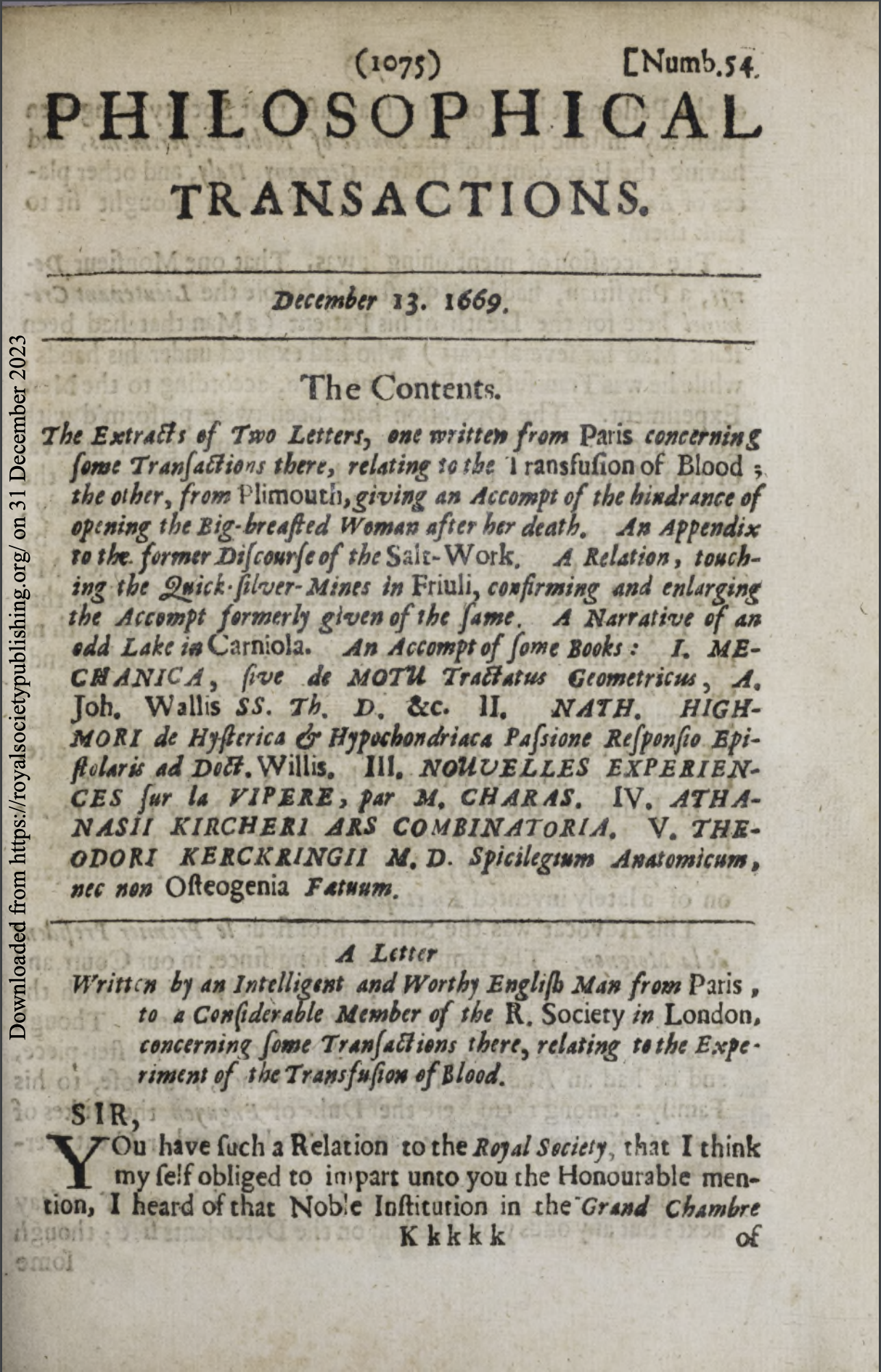}
\caption{A page from the Philosophical Transactions of the Royal Society from December 1669 demonstrating at once the similarity to a modern research article and the significant differences in the level and detail of metadata present in the article \citep{noauthor_letter_1997}.}
\label{F1}
\end{figure}

In the $17^{th}$ Century, at the advent of formalised scholarly communication, there was a relatively small community and communications to the journal either arrived directly from a member of a scholarly society or were submitted via a member.  Thus, names were often either considered immaterial, would have been known from context and are now lost to time, or were elegantly, and seemingly purposefully not mentioned as in the example from 1669 shown in Fig.~\ref{F1}, which reads

\begin{quotation}
\textit{``A Letter Written by an Intelligent and Worthy English Man from Paris, to a Considerable Member of the R. Society in London, concerning some Transactions there, relating to the Experiment of the Transfusion of Blood.''} 
\end{quotation}
The letter is dated but not signed and the author is not explicitly named \citep{noauthor_letter_1997}.

The genesis of any novel form of discourse inherently involves a period of adjustment, wherein norms and social contracts gradually emerge and solidify. This evolutionary process has led to a myriad of reasons for the inclusion or exclusion of specific details in formal communications, some of which mirror broader societal norms. Reflecting on the origins of scholarly communication, primarily rooted in British and European traditions, one can draw parallels with the literary conventions of the time. For instance, in Jane Austen's "Pride and Prejudice" (published in 1797), the first names of Mr. and Mrs. Bennet are never revealed. Similarly, Arthur Conan Doyle's iconic characters, Holmes and Watson, introduced in 1887, seldom use first names in their interactions, reflecting the written norms of their times.

The reasons for omitting author identities can also stem from more problematic motives, such as the desire to conceal an author's gender. Now famously, the Bront\"{e} sisters initially published under male pseudonyms as writing was not considered an appropriate activity for women in that period. More recently J.K.Rowling opted to use her initials to conceal her gender for commercial reasons, anticipating potential bias in her readership, before she became too well known and successful to be able to conceal her identity \citep{whited_ivory_2004}. 

Anonymity or pseudonymity can be a choice not only to protect from persecution, it can also serve to protect authors from repercussions related to the voicing of controversial opinions or work that is not in step with the political regime where it is being carried out. The revolutionary ideas shared on the future of currency are an example in the seminal white paper on Bitcoin by Satoshi Nakamoto, where the identity, gender, nationality, or even the individuality or plurality of the author(s) remains unknown \citep{nakamoto_bitcoin_2008}.

\subsection{The evolution of the research article as technology}

The role of technology in facilitating or restricting the disclosure of author information has pivotal import. For example, while current practices do not mandate the encoding of gender data in author identities, this omission has been significant in fields where research outcomes vary with the researcher's gender \citep{sorge_olfactory_2014,katsnelson_male_2014,reardon_sex_2017}. The decision not to maintain detailed records of authors' gender or ethnicity is influenced by contemporary societal norms, yet future scholars might view this lack of information as a critical oversight, akin to how the absence of affiliation details in early scholarly works is perceived today.

Perhaps more importantly than the needs of future scholars, the increase in transparency afforded by first author names plays a daily role in how we experience research.  First names, in the ethnicities and genders that they suggest, provide an, albeit imperfect, high-level reflection of the diversity of experiences that are brought to research. It is just as important to see ourselves reflected in the outputs of the research careers that we choose to pursue as in the voices that represent us on panels at conferences; recent work suggests that demographic representation in the scholarly record is materially linked to the diversity of innovation that science produces \citep{hofstra_diversity_2020}.  This highlights the non-neutral role of technology in the representation of individual-related information.

\begin{figure}[h!]
\centering
\includegraphics[width=.95\linewidth]{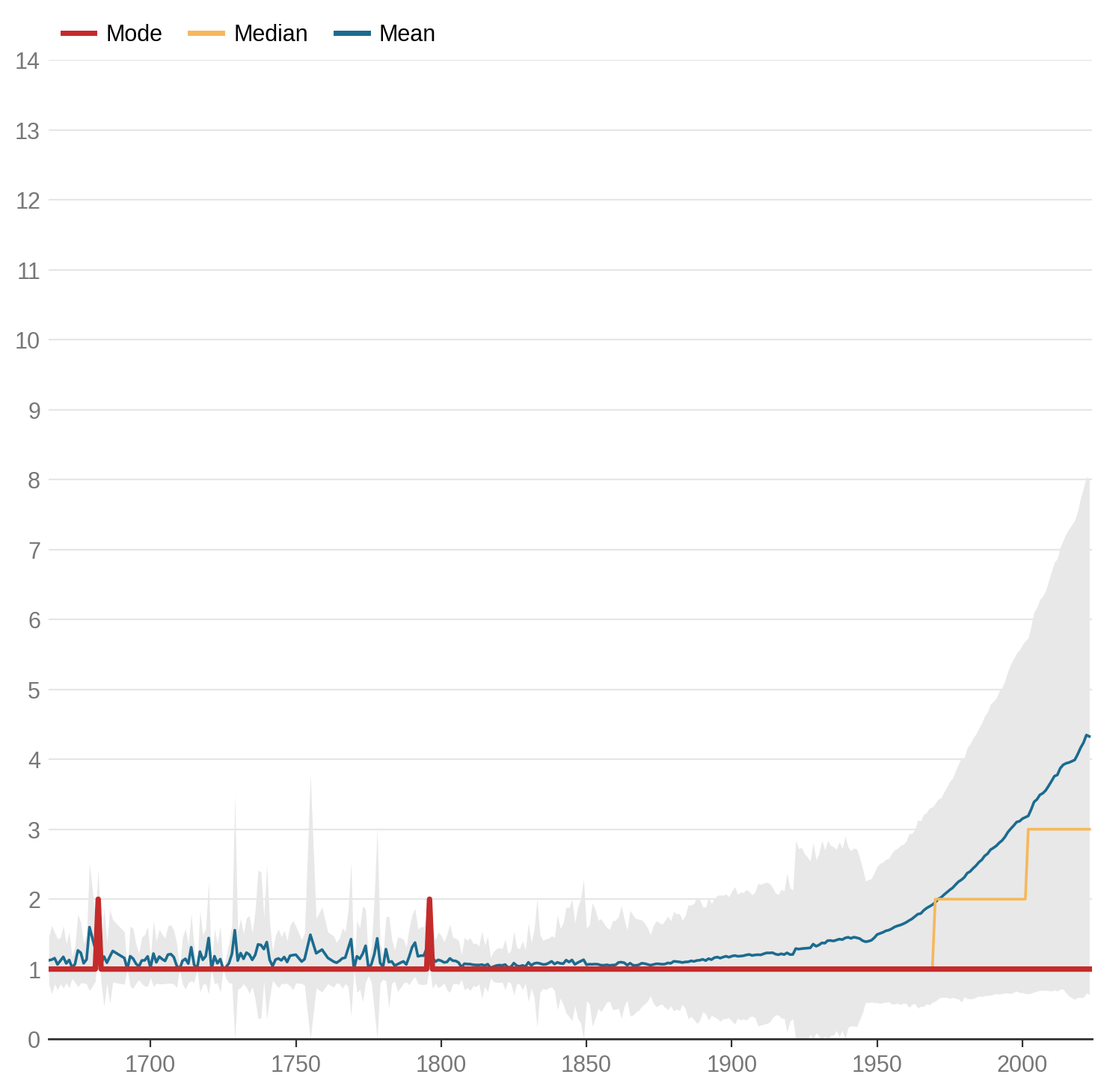}
\caption{Development of averages of co-authorship numbers on academic output since 1665.  The red line shows the modal number of co-authors, the yellow line shows the median number of co-authors, the blue line shows the mean average number of co-authors.  The grey shaded area shows the standard deviation from the mean bounded below by the axis. Each article is counted once.}
\label{F2}
\end{figure}

The interplay between technological and societal changes, each influencing the other in complex ways, is also evident in the shifting patterns of co-authorship in scholarly papers. Historical analysis (see Fig.~\ref{F2}) reveals that the median number of authors on an article was one until 1970, two until 2002 and three thereafter, while the mean rose from 1.4 in 1945 to 4.3 in 2023, reflecting the changing sociology of research.  Analysis of the locations of researchers shows the fundamental nature of research community \citep{adams_fourth_2013}. 

Three insights frame the analysis that follows:

\begin{enumerate}

\item \textbf{Gradual Evolution of the Scholarly Record}: The development of conventions around the scholarly record, such as the norm of naming authors on papers, reveals a slow evolutionary process. It highlights that it took nearly two centuries to establish a general expectation for author attribution on scholarly works. This gradual change underscores the inertia inherent in academic traditions and the time it takes for new norms to solidify across the research community.

\item \textbf{Accelerated Changes in Recent Times}: The pace at which the scholarly record has been transforming has markedly increased in recent years. This acceleration is evident in the tripling of the mean number of authors per article, from 1.4 to 4.3, in the 80 years since 1945, a change driven by the increasing frequency, volume, and data size of scholarly outputs. This contrasts sharply with the nearly 300-year period it previously took for such a demographic shift in authorship, highlighting how technological advancements and the expanding scope of collaborative research have expedited changes in scholarly communication practices.

\item \textbf{Reflective Richness of Data}: The data surrounding scholarly communication not only mirrors the evolution of the research community but has, in recent times, also become indicative of the technologies that facilitate this communication. The growing complexity and interconnectedness of research outputs, evidenced by the increasing number of co-authors and the expansion of collaborative networks, reflect broader shifts in the technologies available for research and dissemination. This richness in data offers profound insights into the dynamics of scholarly practices and their evolution over time.
\end{enumerate}

The slow and steady evolution shown in Fig.~\ref{F2} in collaboration may also be considered to be a societal parallel for the slow evolution in the nature of the technologies that underpin scholarly communication. The persistence of the single-authored article as the most common form from the genesis of scholarly communication to the present day, set against a mean that has tripled since 1945 and a median that has moved from one author to three, gives us a sense of when technological and social changes started to facilitate and change norms. A more detailed analysis of evolution of co-authorship and its importance in addressing more complex problems is dealt with by Thelwall and Maflahi \citeyearpar{thelwall_research_2022}.

The 20$^{\rm th}$ century witnessed a transformative shift in research practices, fundamentally driven by technological advancements. This period marked the emergence of what Weinberg \citeyearpar{weinberg_impact_1961} termed ``Big Science'', a paradigm characterised by large-scale scientific endeavours that required extensive collaboration across disciplines and often, across national borders, and which de Solla Price \citeyearpar{desolla_little_1963} subsequently analysed as a quantitative phenomenon laying the foundations of modern scientometrics. The post-war expansion was accompanied by a substantial increase in government R\&D spending, particularly in the United States following Sputnik, and by the emergence of the postdoctoral research position as a routine career stage \citep{stephan_economics_2012}---both developments that fundamentally restructured the demography of research. The sociological landscape of certain research fields underwent significant changes due to this shift, influencing and reshaping established norms within the scientific community. The impact of Big Science is particularly evident in the evolving patterns of authorship, as collaborations expanded to include hundreds or even thousands of contributors.

\begin{figure}[h!]
\centering
\includegraphics[width=.95\linewidth]{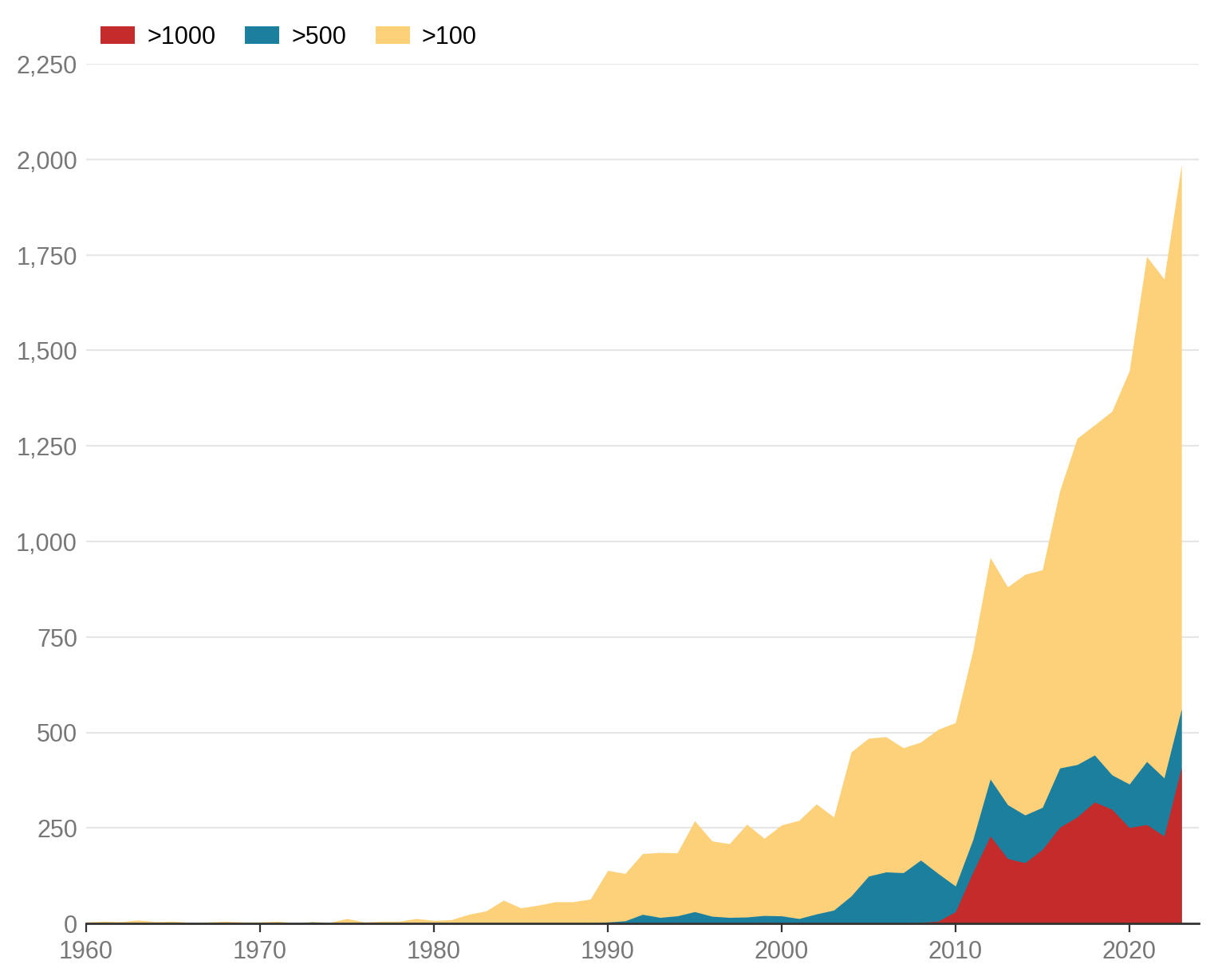}
\caption{Number of papers in each year with more than 100 (yellow area), 500 (blue area) and 1000 (red area) co-authors respectively. These are layered, not stacked, series: each is plotted independently from a common zero baseline, drawn back-to-front from the largest ($>100$) to the smallest ($>1000$) category, so the visible bands show each threshold's actual annual count rather than an additive total.}
\label{F3}
\end{figure}

Illustrative of this trend, the growth in the number of research papers featuring extensive co-authorship is striking (see Fig.~\ref{F3}). Papers with more than 100 authors, indicative of large collaborative projects, began to appear with greater frequency, reflecting the complex, interdisciplinary nature of modern scientific inquiries (yellow region) starts in earnest in the early 1980s \footnote{The first paper in \textit{Dimensions} with co-authorship of more than 100 authors is in Chemistry from 1928 \citep{emich_allgemeine_1928}.}. This trend is further accentuated in research requiring substantial resources, such as particle physics experiments, where the collective expertise and effort of hundreds of scientists are essential to the project's success. Papers with over 500 co-authors (blue) and, notably, more than 1000 co-authors (red) have become increasingly common, underscoring the scale of collaboration in certain areas of research.

On a practical level, the significance of author names extends beyond mere attribution. In the context of evaluation for promotions, tenure, grant assessments, and many other forms academic recognition, the identity of authors, corroborated by associated identifiers, plays a crucial role. Indeed, research is becoming increasingly centred around quantification \citep{pardo-guerra_quantified_2022}. The trust placed in a particular paper, and by extension, in communities of researchers, is increasingly dependent on the clarity and reliability of authorship information. Furthermore, this information is pivotal in selecting referees, assessing potential conflicts of interest, and facilitating a myriad of other critical academic processes. The evolution of scholarly communication, therefore, not only reflects changes in the academic landscape but also emphasises the growing importance of clear and reliable attribution in maintaining the integrity and trustworthiness of scholarly work.

\section{Methodology}
In the study presented here, we explore the change in the representation of author names in scholarly research articles.  As established in the Introduction, we define the terms \textit{initial form} to refer to an author name where only the initials are supplied and no given name is supplied, and \textit{full form} to refer to an author name that includes at least one fully stated given name. We use \textit{Dimensions} from Digital Science as the data source for our exploration. At the time of writing \textit{Dimensions} contains over 144 million publications, of which approximately 112 million are journal articles.  Much of the analysis shown in this paper is based on queries that can be run directly on the \textit{Dimensions} database in the Google BigQuery environment without the need for further coding in, for example, a Python environment and hence are accessible to a broad audience \citep{hook_scaling_2021, porter_connecting_2022}.

\subsection{The layered provenance of bibliographic data}
\label{sec:provenance}
Before describing our analytical approach, it is important to acknowledge that \textit{Dimensions}, like all major bibliographic databases, is a layered artefact whose coverage and metadata fidelity reflect the accumulated history of the systems it ingests. The modern bibliographic record is a composite of multiple data streams established at different times for different purposes: MedLine, with retrospective coverage to 1946 and rolled out as an electronic system in the 1960s; PubMed, launched in 1996 with a substantial platform redesign in 2002 that introduced full given-name metadata fields; PubMedCentral, launched in 2000; the Crossref DOI infrastructure, whose metadata schema accommodated given names from inception; and JATS XML feeds from publishers, increasingly comprehensive through the 2000s and 2010s. Each system carries its own coverage discontinuities, curation policies, and metadata conventions, and these features propagate into any database that ingests them.

Some of the discontinuities we observe in the data are therefore not features of scholarly practice itself but of the systems that record it. We engage with this directly in Section~\ref{sec:tech} (Technological Analysis), where the contrast between PubMed-with-DOI and PubMed-without-DOI subsets allows us to separate metadata-driven from practice-driven changes. As we argue in the Discussion, this distinction is itself central to the contribution of bibliometric archaeology as a methodology.

\begin{figure}[h!]
\centering
\includegraphics[width=.95\linewidth]{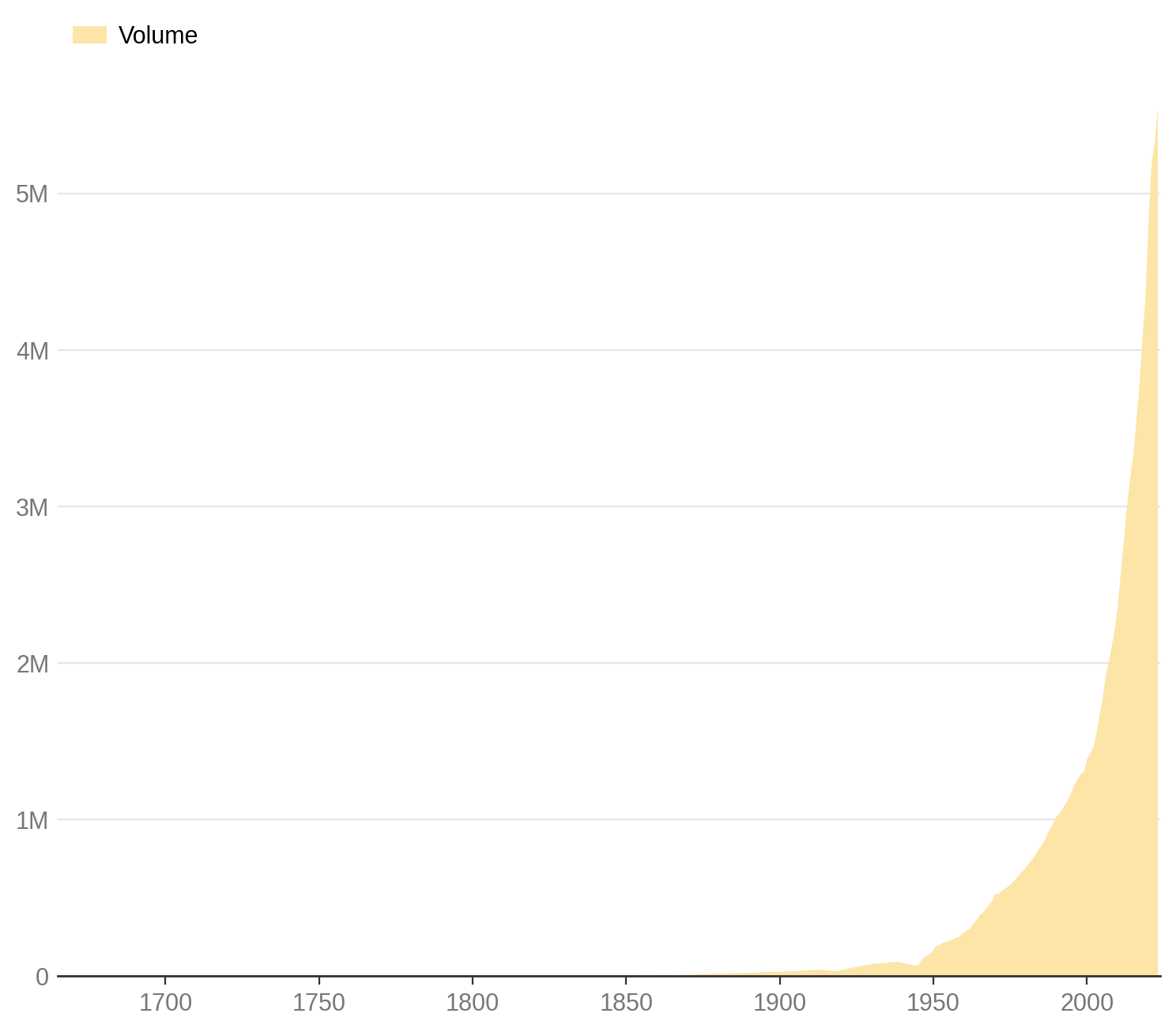}
\caption{Volume of scholarly publication using the restrictions specified in Listing~1 for comparability with other plots in the paper.  This plot shows a clear cutoff around 1940 at which point there is sufficient data for the reasonable interpretation of statistical approaches.}
\label{F4}
\end{figure}

Although we have included several graphs and examples regarding scholarly output before 1945 we recognise that publications volumes are smaller further back in time.  Thus, we define two epochs in publication, one before 1945 which we refer to as the \textit{anecdotal epoch} and the other after 1945 that we call the \textit{statistical epoch}.  It is noteworthy that \textit{Dimensions}, our primary data source, is limited in the validity of its data for earlier times in the anecdotal epoch.  The data source was not constructed with the current use case in mind and hence coverage of content in the 17$^{\rm th}$, 18$^{\rm th}$ and even 19$^{\rm th}$ centuries is reliant on publisher and community interest in awarding persistent identifiers to research outputs of these periods in order for them to be included in \textit{Dimensions}.  In addition, mappings to institutions that have disappeared or changed their name over the earlier part of the last 350 years may not be fully represented in geographical analyses - similarly, country references will be to modern countries not countries that have existed in the past - we take no account of moving boundaries, merely imposing the current world view back in time for simplicity of analysis. Fig.~\ref{F4} shows the number of articles available in each year; prior to approximately 1945 the annual counts are small enough that we treat the data as anecdotal, and this is one reason for our reluctance to perform statistical inference in that epoch.

Broadly speaking we have set out not to use statistical methods or to take inferences based on data in the anecdotal epoch - such figures that depict these data are intended to be illustrative, sometimes highlighting the limitations of \textit{Dimensions} for the period, and at other times to add anecdotal colour and context to our discussions and arguments. Data concerning the statistical epoch, is more trustworthy both from a structural perspective (systems existed to more completely capture the scholarly record and it is a period during which there remains significant academic interest in reading and making reference to the material, thus the literature of this period has much better coverage of persistent identifiers) as well as from a volume perspective. Fig.~\ref{F4} shows the growth in volume of publications (using the restrictions specified in Listing~1).

In terms of both the quality and volume of the data, the transition from the anecdotal to the statistical epoch is not sharply defined but happens gradually.  However, our choice of 1945 is not completely arbitrary.  The current authors have previously noted that 1945 was the year in which the centre of mass of global research stopped its march toward North America and began to move back toward Europe \citep{hook_scaling_2021}.  It was also the year in which Vannevar Bush published the Endless Frontier \citep{bush_endless_1945} and, as such, is a good point in time from which to date the modern approach to research. Indeed, both Figs~\ref{F4} and~\ref{F5} show a sharp change in behaviour precisely at 1945, not a gradual drift across the late 1940s: the global research enterprise was reset in that single year by the combined effect of the end of the war in Europe and the UK---where research infrastructure had been disrupted and began to be rebuilt in the image of what we might call the ``Vannevar Bush Era'', an era defined by the Endless Frontier \citep{bush_endless_1945}.

On a technical note, during the \textit{Dimensions} data ingestion process author details are gathered from across the available sources that \textit{Dimensions} draws on. Each author has a last name and a first name, a unique identifier assigned within \textit{Dimensions} (where it is possible to determine), a list of institutional affiliations mapped to unique identifiers (GRIDs, where resolvable), an ORCID (again, where resolvable) that have been asserted in relation to a given output, and a marker of whether the author is listed as a corresponding author.  For the majority of publishers this is delivered via a JATS XML feed, but for smaller publishers or those without the technical capability to deliver JATS XML, Crossref data is used or, with the permission of the publisher, data is crawled directly from their website.  Once cleaned and enhanced in the \textit{Dimensions} data pipeline, the data are parsed into the \textit{Dimensions} database and loaded into Google BigQuery.

Each analysis that we present takes either an author-centric or paper-centric view.  That is to say we either examine the proportion of papers published in a year with certain characteristics or we examine the characteristics of authors that have published in the year directly.  As a side note, we observe that the active publishing community is a reasonable proxy for the global research population but at each point in the timeline that we engage with, there are good reasons why this cannot be taken to be more than a representative sample.  For example, in the early years of the scholarly record (circa 1665) publication was a new activity that was not engaged with by every academic and social structures of the time tended to exclude women.  On the other hand, in the present day, a significant proportion of global research takes place in proprietary settings and hence is not published in the scholarly record.  Thus, all the comments that we make in the paper need to be taken within these constraints in mind.

\begin{lstlisting}[language=SQL,caption={\label{lst:1}Core SQL for \textit{Dimensions} on Google BigQuery: the given-name classifier and the author-position table from which every analysis in this paper is derived (deposited as \texttt{fig00\_base\_tables} in the accompanying code).},captionpos=b]
-- Given-name classifier (BigQuery SQL functions)
CREATE TEMP FUNCTION name_tokens(fn STRING) RETURNS ARRAY<STRING> AS (
  ARRAY(SELECT t FROM UNNEST(SPLIT(REGEXP_REPLACE(TRIM(IFNULL(fn,'')),
        r'[\.\-\s]+', ' '), ' ')) t WHERE t != ''));
CREATE TEMP FUNCTION has_cjk(fn STRING) RETURNS BOOL AS (
  REGEXP_CONTAINS(IFNULL(fn,''), r'[\p{Han}\p{Hiragana}\p{Katakana}\p{Hangul}]'));
CREATE TEMP FUNCTION classify_given_name(fn STRING) RETURNS STRING AS (
  CASE WHEN fn IS NULL OR TRIM(fn) = ''
         OR ARRAY_LENGTH(name_tokens(fn)) = 0       THEN 'undetermined'
       WHEN has_cjk(fn)                              THEN 'undetermined'
       WHEN (SELECT LOGICAL_AND(LENGTH(t) = 1
                     OR REGEXP_CONTAINS(t, r'^[A-Z]{2,3}$'))
             FROM UNNEST(name_tokens(fn)) t)         THEN 'initial'
       ELSE 'full' END);
-- One row per author position, classified BEFORE affiliations are expanded
SELECT p.id AS pub_id, p.year, pos AS author_pos,
       ARRAY_LENGTH(p.authors) AS n_authors,
       author.first_name, author.last_name, author.researcher_id,
       classify_given_name(author.first_name) AS name_class,
       (SELECT aff.country_code
          FROM UNNEST(author.affiliations_address) aff WITH OFFSET aff_pos
         WHERE aff.country_code IS NOT NULL
         ORDER BY aff_pos LIMIT 1)             AS primary_country_code
FROM `dimensions-ai.data_analytics.publications` p,
     UNNEST(p.authors) AS author WITH OFFSET AS pos
WHERE p.type = 'article'
  AND ARRAY_LENGTH(p.authors) BETWEEN 1 AND 49
  AND p.year BETWEEN 1665 AND 2023
\end{lstlisting}

It would be a simple matter to test the first name to see if it is of length one or two characters (either an initial or an initial and a ``.'').  However, this approach would miss people with multiple initials or with a two character first name such as ``Bo''.  We therefore split the given-name field on spaces, full stops and hyphens and examine the resulting tokens (Listing~1).  A name is classified as \textit{initial form} if every token is a single character or an undotted run of two or three capital letters (the legacy MEDLINE representation of initials, e.g. ``JR''); as \textit{full form} if any token has two or more characters otherwise (``Bo'', ``Daniel W'', and the case of authors known principally by their second name, such as J. Robert Oppenheimer, see for example \citealp{oppenheimer_communication_1963}); and as \textit{undetermined} if the field is blank or contains characters from a script in which a single character is a whole name (see below).

A further complexity is that, in many cases, names are preserved in both their original and transliterated forms in \textit{Dimensions}.  In Chinese, Japanese and Korean scripts a whole given name may be a single character (see, for example \citealp{yan_clinical_2005}) and would be mistaken for an initial by a length-based test.  Rather than treat such names as initials, we place any given-name field containing Han, Hiragana, Katakana or Hangul characters in the undetermined category.  The effect is modest: as of the date of writing, only 250,991 of 8,456,140 publications in Chinese (around 3\%) and 1,152,602 of 4,460,000 Japanese language publications (around 25\%) have the potential to contain such characters, and undetermined names account for between 2\% and 6\% of papers in every year after 1945 (Fig.~\ref{F5}).

We have limited the query in Listing~1 to include only outputs that are classed as papers, since books and conference proceedings have different qualities and characteristics that we don't wish to cloud the current discussion.  We have also removed papers with more than 50 co-authors as these papers also have different characteristics.

We have built further aggregations on top of the core table provided by Listing~1 to support the other analyses presented in this paper.  Once the data are presented in the format from Listing~1 we perform aggregations that give a determination of the names of authors at both an author and a paper level.  Each author position on a paper is classified as initial form, full form or undetermined as described above.  A paper is then classified as initial form when every classifiable author is in initial form, full form when every classifiable author is in full form, ``Mixed'' when both forms are present, and ``Undetermined'' when no author on the paper can be classified.  The structure of Listing~1 allows us then to study how the data change with time, with subject category and with country.

In exploring the issue of gender we used genderize.io under licence.  The genderize.io tool provides a dataset with probabilities that a given name is statistically associated with either  men or women, or ``undetermined'' on both a nation-neutral and national basis.  The tool is well-used in the study of gender in academia \citep{bruck_bibliometric_2023, holmes_gender_2019}.  This is a statistical analysis and hence cannot attempt to access the subtle and complex issues of gender identity, but rather it focuses on how gender is represented in the historical scholarly record.  We have employed a methodology where we matched name variations in author names in \textit{Dimensions} to the results obtained for that name variation in genderize.io. The results were stored in a private table in Google BigQuery. We then matched author names, where given, to data in this table.  We used the country of the author's first-listed affiliation on the paper to select the country-specific genderize.io result where one existed, and the nation-neutral result otherwise.  The genderize.io data are used under licence and are not redistributed with the code; the lookup procedure, source date and fallback rule are documented in the deposited README.  Obviously, this approach is open to errors in that, especially in more recent years, researchers are much more likely to travel during their careers and hence their current university may not be culturally relevant for such an analysis.  Many names are also used much more internationally now and hence those that are traditionally nationally associated may not be used in context.  It is also the case that some researchers will originate from countries where a Western name is adopted for business purposes.  In this final case, the names are often more central to common usage as they are designed to mimic Western-style naming conventions and may even improve the data.  The point of this paper is not, however, to do a detailed gender analysis and we believe that our approach is sufficiently robust to be serviceable in the current context. 

\section{Results}
\subsection{Paper-based analysis}
\begin{figure*}[th!]
\centering
\includegraphics[width=.95\linewidth]{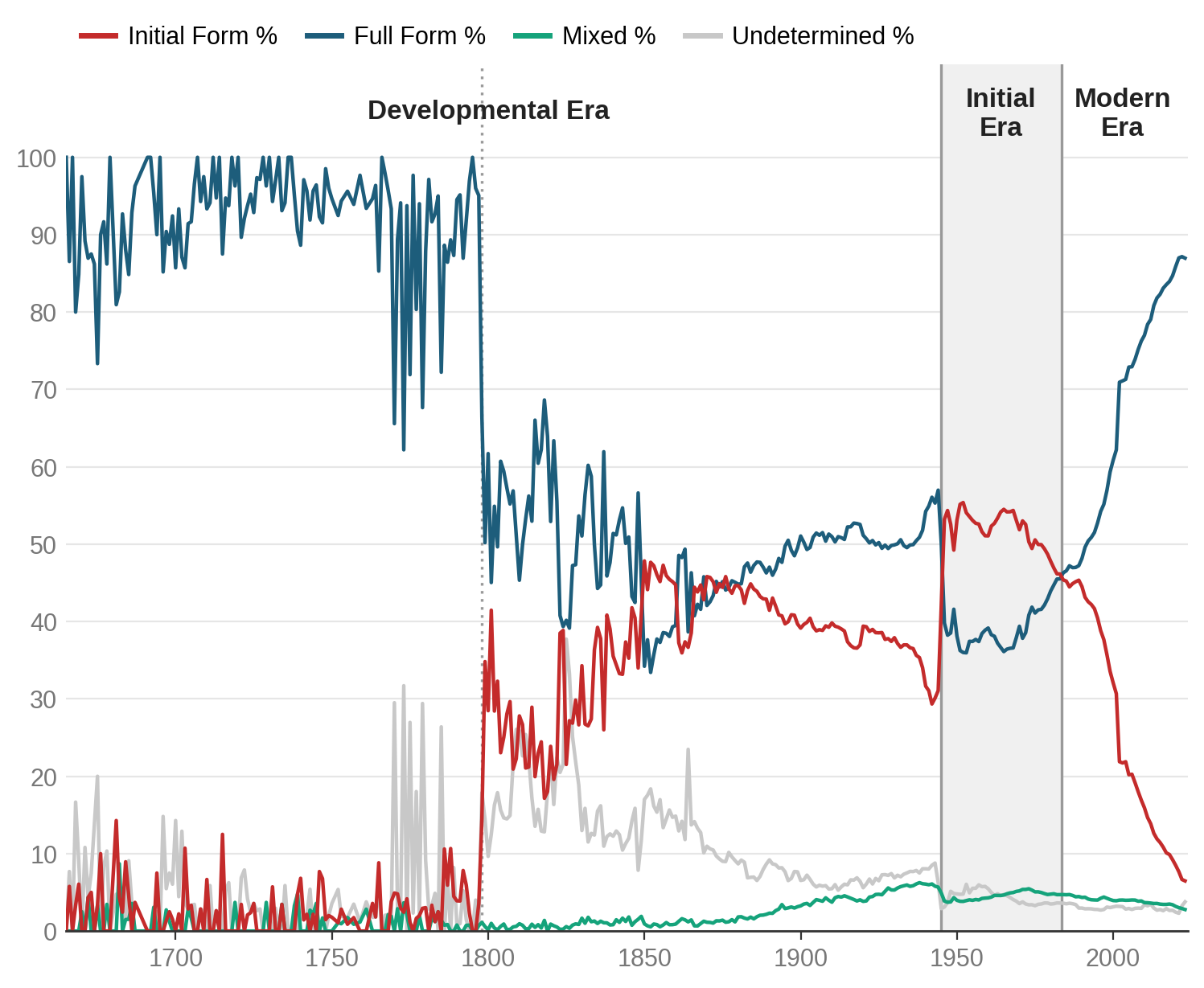}
\caption{Proportion of papers in which all authors state their names using initials (``Initial form \%'' - red line), versus all authors stating their full names (``Full form \%'' - blue line) versus some authors stating their full name while others state their initials (``Mixed \%'' - green line.  ``Undetermined \%'' comprises papers on which no author can be classified, either because no given name is recorded or because the given name is in a Chinese, Japanese or Korean script.  We note three distinct eras of behaviour: The Developmental Era tracks from the genesis of scholarly publication until 1945; The Initial Era from 1945 to 1983, the shaded band ending where the initial-form and full-form lines cross; and the Modern Era from 1983 to present day.}
\label{F5}
\end{figure*}

As indicated in the introduction, the historical use of names in the scholarly record has been uneven and may, to a certain extent, reflect societal norms.  While, in the early years of \textit{Dimensions}' coverage, the volumes of data are small, Fig.~\ref{F5} shows that until around 1800 it appeared to be normal to use one's full name in scholarly correspondence (despite the fascinating cases discussed in the introduction).  It is also important to acknowledge the lens through which we are examining the past is imperfect.  The work of Lefanu \citeyearpar{lefanu_british_1937} makes it evident that there was a lively and thriving research discussion prior to 1800, however, records have either not been preserved or not been transferred into digital realms, meaning that they are invisible to our study.

\begin{figure}[h!]
\centering
\includegraphics[width=.95\linewidth]{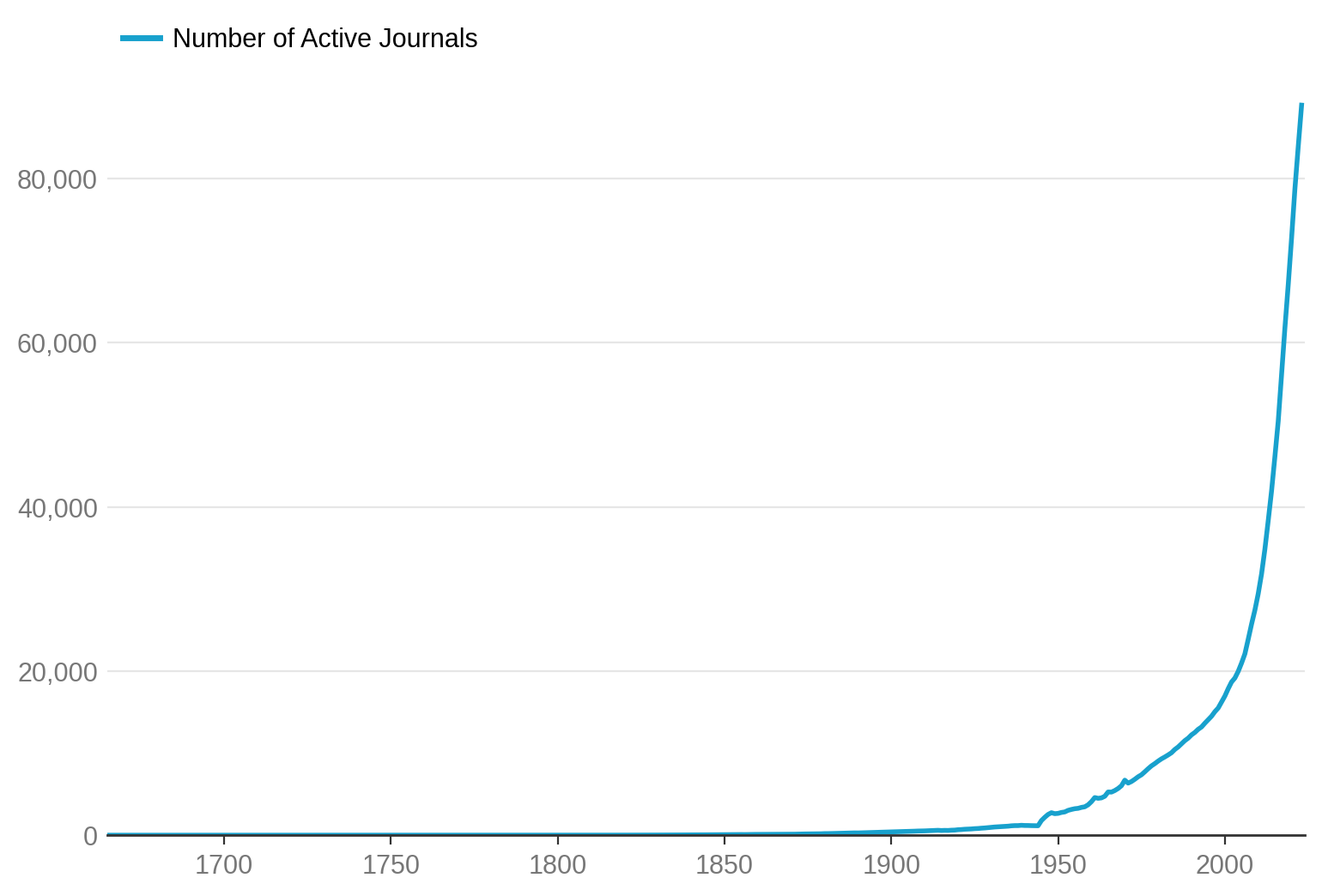}
\caption{Development in the number of actively publishing journals.}
\label{F6}
\end{figure}

Figure~\ref{F5} can be broken into three eras, with one phase aligned with the anecdotal epoch, and the second and third phases in the statistical epoch:
\begin{itemize}
    \item We define the ``Developmental Era'' as the period from 1665 to 1945, that appears to be split into two sections, separated by a sharp transition that occurs in 1798.  As we will explain, the sharpness of this distinction is a result of the data in \textit{Dimensions} (and of the systems on which \textit{Dimensions} is based--specifically, systems that are based around the attribution of DOIs to academic articles \citep{hook_dimensions_2018}).  Thus, we must regard the data reported until 1798 as unrepresentative of the actual development of scholarly publication and only anecdotal in nature.
    
    In the period until 1798 the Dimensions data contains just a few regularly publishing journals with total global number of articles being published reaching 179 in 1798.  Up until that year, the journals with the largest publication volumes were Philosophical Transactions of the Royal Society of London, Transactions of the Linnean Society of London and Transactions of the Royal Society of Edinburgh, each of which were publishing around 30 articles per year and all of which had adopted a standardised title page, typically printed the names of authors in full form.  But, in 1798, \textit{The Philosophical Magazine: A Journal of Theoretical, Experimental and Applied Physics} was established by Alexander Tilloch of the London Philosophical Society, publishing 142 articles; the following year, in 1799, an even more voluminous medical journal appears, recorded in \textit{Dimensions} under the title of its successor, \textit{The British and Foreign Medical Review} (a title that was itself not founded until 1836), publishing 327 articles. Hence, there was a sharp and significant increase in the number of articles in \textit{Dimensions}. 
    These are the articles in publications that scholarly societies and other participants in the scholarly ecosystem decided were sufficiently valuable (and where means were sufficient) to clean up metadata and make DOIs or PubMed identifiers available. In the two significant journals mentioned here, author name forms are mixed in both journals with some articles being editorial (without author names), but while full form names still appear regularly, initial form becomes much more acceptable and this becomes established as a dominant signal entering the 1800s. Lefanu provides a for more detailed treatment of the rise of medical journals \citeyearpar{lefanu_british_1937}.
    
    In the years following 1800 the numbers of articles increase in volume to a level where some types of statistical analysis are appropriate, so long as the data are not filtered and segmented too finely.  By 1826, the total, global number of journal articles tracked in \textit{Dimensions} is consistently above 1000 (the level of a mid-sized research university today).
    
    It is worthy of note that throughout the whole of the Developmental Era there a low level of co-authorship and hence, we remain in Adams' first age \citep{adams_fourth_2013}. As we see in Fig.~\ref{F2} the median number of authors on a paper does not reach two until 1970.
    
    During this period scholarly communications are published in early versions of scholarly journals operated by early scholarly societies, with an intended audience of members, almost as we would treat newsletters today.  Personal connections and members would often have been known to each other due to the small size of these early communities.  Several innovations took place in this period, however, including the introduction of a somewhat standard title page to communications.

    In 1798, at the transition point, only 179 papers were published (as tracked by \textit{Dimensions} under the constraint of Listing~1), but in 1799, 514 papers were published - almost a tripling of output. This significant increase in output appears to be coupled to the beginning of the proliferation of scientific journals.  Although it is difficult to see the detail of this development in Fig.~\ref{F6}, which shows that there were around 89,000 actively publishing journals in 2023, in 1797 just five journals are recorded as active in publication in \textit{Dimensions} (until the mid-1780s, more than a century after the beginning of this movement, there had only been one or two active regularly publishing journals at any point in time, with 1786 being the first year when three journals published). From 1799, eight journals were regularly publishing.  By 1827 this had become 19 and by 1837 it was 33.  This part of significant growth began to see the development of norms and standards that needed to reach across journals and the community started to become large enough that the prior ``gentlemanly'' approach was no longer sufficient to handle the level of publication volumes and the community which needed to share in the knowledge being generated.  
    
    During this period, the level of formality in the use of initial form versus full form in papers appears to have increased spontaneously and significantly.
    
    \item The ``Initial Era'' began in 1945, at the point where the disruption and reconstruction of research in the UK and Europe coincided with the US's demobilisation from war and the dawn of Bush's Endless Frontier \citep{bush_endless_1945}, and continues until 1983, the last year in which initial-form papers outnumber full-form papers (Fig.~\ref{F5}).  It is the singular period in the history of the scholarly record in which the initial form dominated the full form of author names. The initial era is entirely contained in the statistical epoch and hence there are sufficiently many journals and papers that a statistical approach to analysis is valid.

    \item The ``Modern Era'' has persisted from 1983 to the present day.  It is an era in which we have seen the norm move steadily and rapidly toward the dominance of the use of full form names.
\end{itemize}

\subsection{Researcher-based analysis}
In Fig.~\ref{F5} we see the proportion of papers in which authors used initial form.  In this section we ask a slightly different but related question, which is: How many authors used initial form?  While one might expect this to be an identical question, and while it is related, there is a fundamental difference.  Papers as objects exist in a world of journals, subject norms, publisher house styles and technological constraints.  While each of these affects the authors, they have strong ties to national, cultural and linguistic effects.  Thus, analyses of people rather than papers do add to our understanding of the developments shown above.

\begin{figure}[h!]
\centering
\includegraphics[width=.95\linewidth]{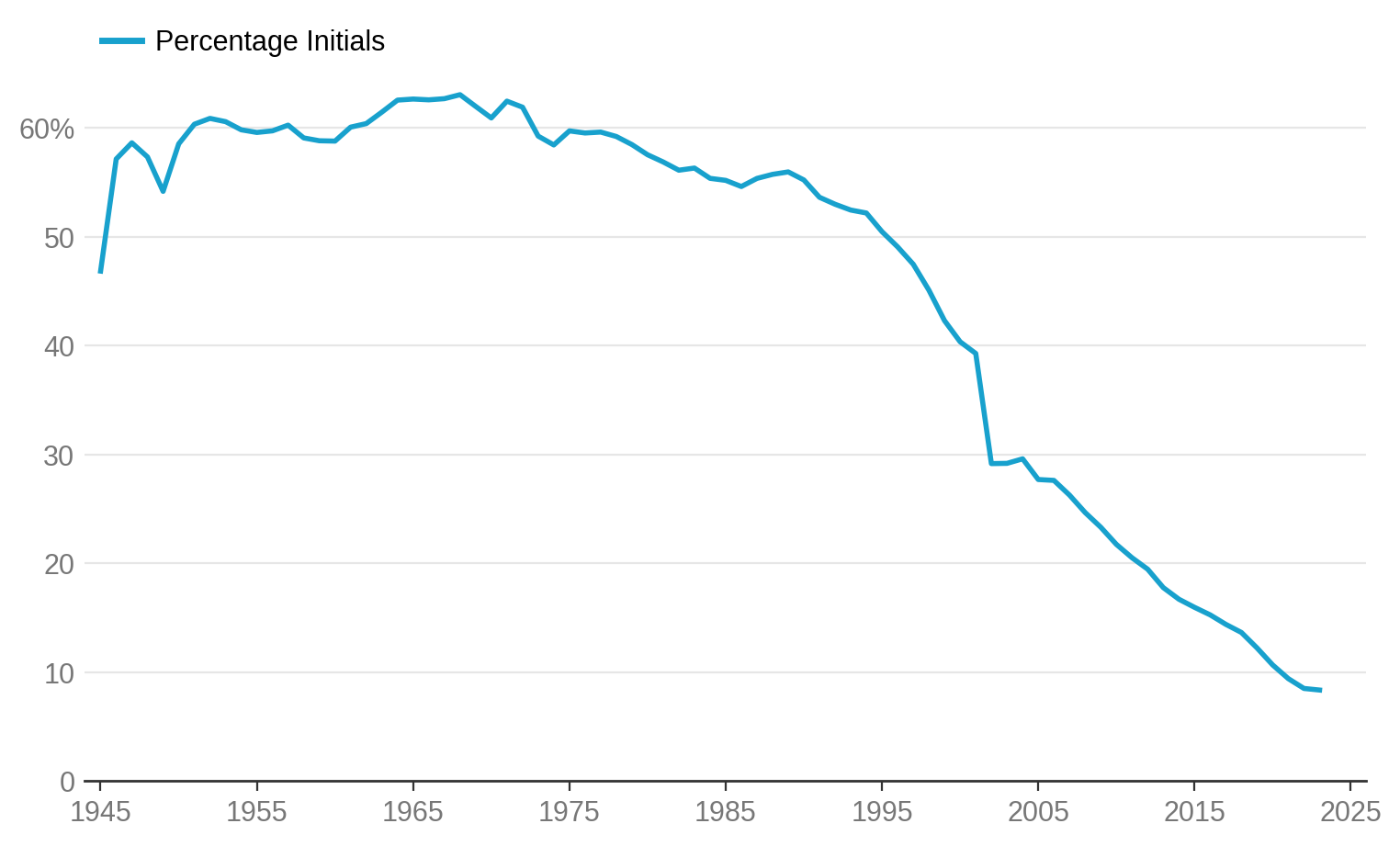}
\caption{Proportion of author positions (one per author per paper) in initial form among those that can be classified, 1945 to 2023.}
\label{F7}
\end{figure}

Figure~\ref{F7} shows the globally conglomerated author view of the use of initial form on journal articles---it averages across national, cultural, gender, technological and disciplinary boundaries.  We will decompose this plot in a variety of ways through the rest of the Results section, so it is important to gain familiarity with its overall shape and its key features.  It is worthy of note that there is a an initial steep incline from 1945, marking the beginning of the Initial Era.  The norm of initial-form dominance is maintained between 55\% and 63\%, peaking in 1968, until the late 1970s, when the form starts being replaced by full form names.  There is a steady decrease in popularity of initial form between 1990 and 2000 (from 55\% to 40\%) and then, in 2002, a precipitous drop from 39\% to 29\%, followed by a continued decline to below 10\% (8.4\% in 2023).  It looks likely that the current decline will continue until initial-form is expunged entirely (or almost entirely).

Had we plotted Fig.~\ref{F7} into the past, we would see that long-form names not only dominated the paper-based view but also the researcher-based view of publication in the anecdotal epoch.

\subsection{Geographical analysis}
\label{sec:geo}
\begin{figure}[h!]
\centering
\includegraphics[width=.95\linewidth]{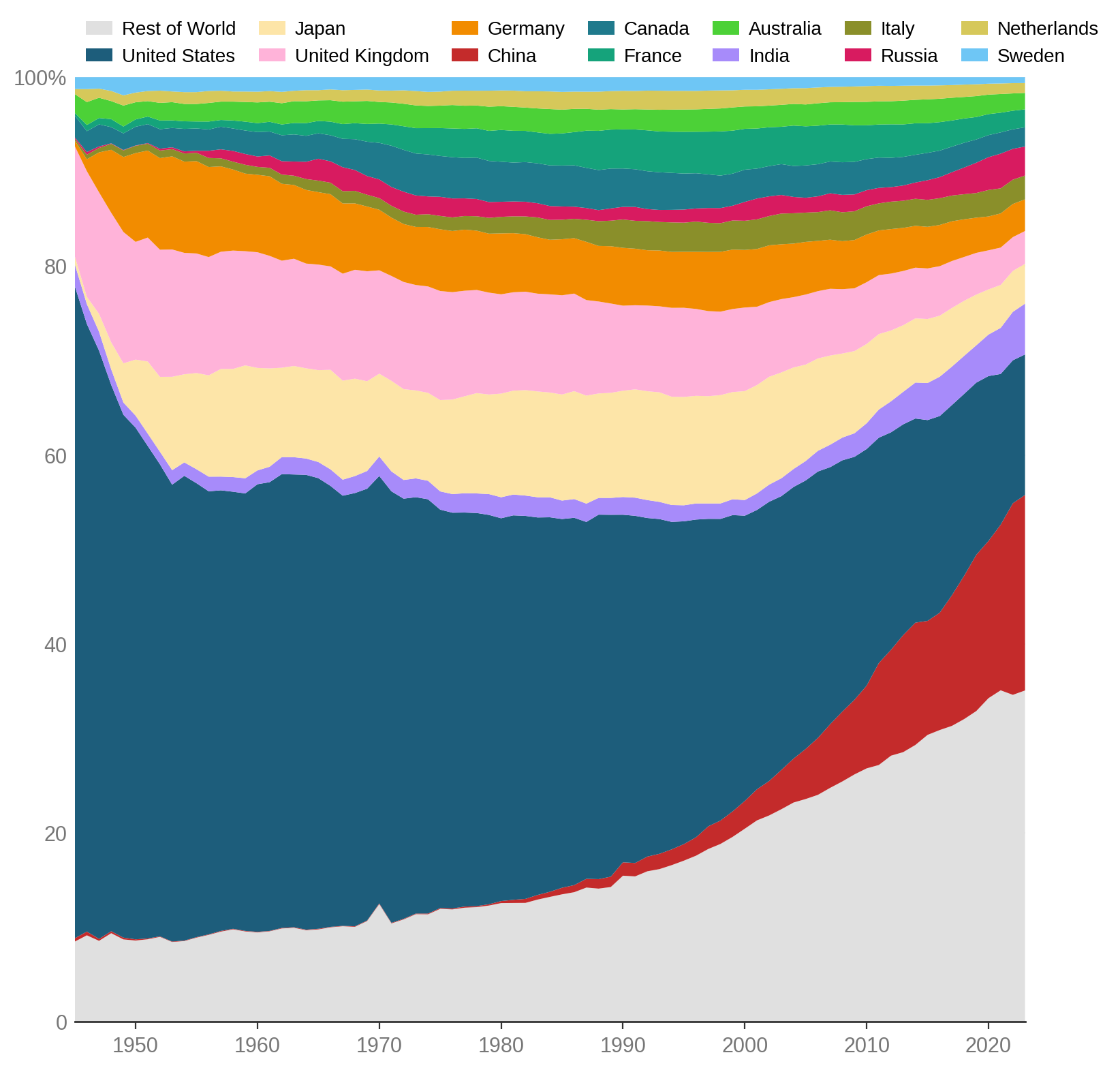}
\caption{Percentage annual contribution to research journal publications by country of affiliation of author from 1945 to 2023.}
\label{F8}
\end{figure}

\begin{figure}[h!]
\centering
\includegraphics[width=.95\linewidth]{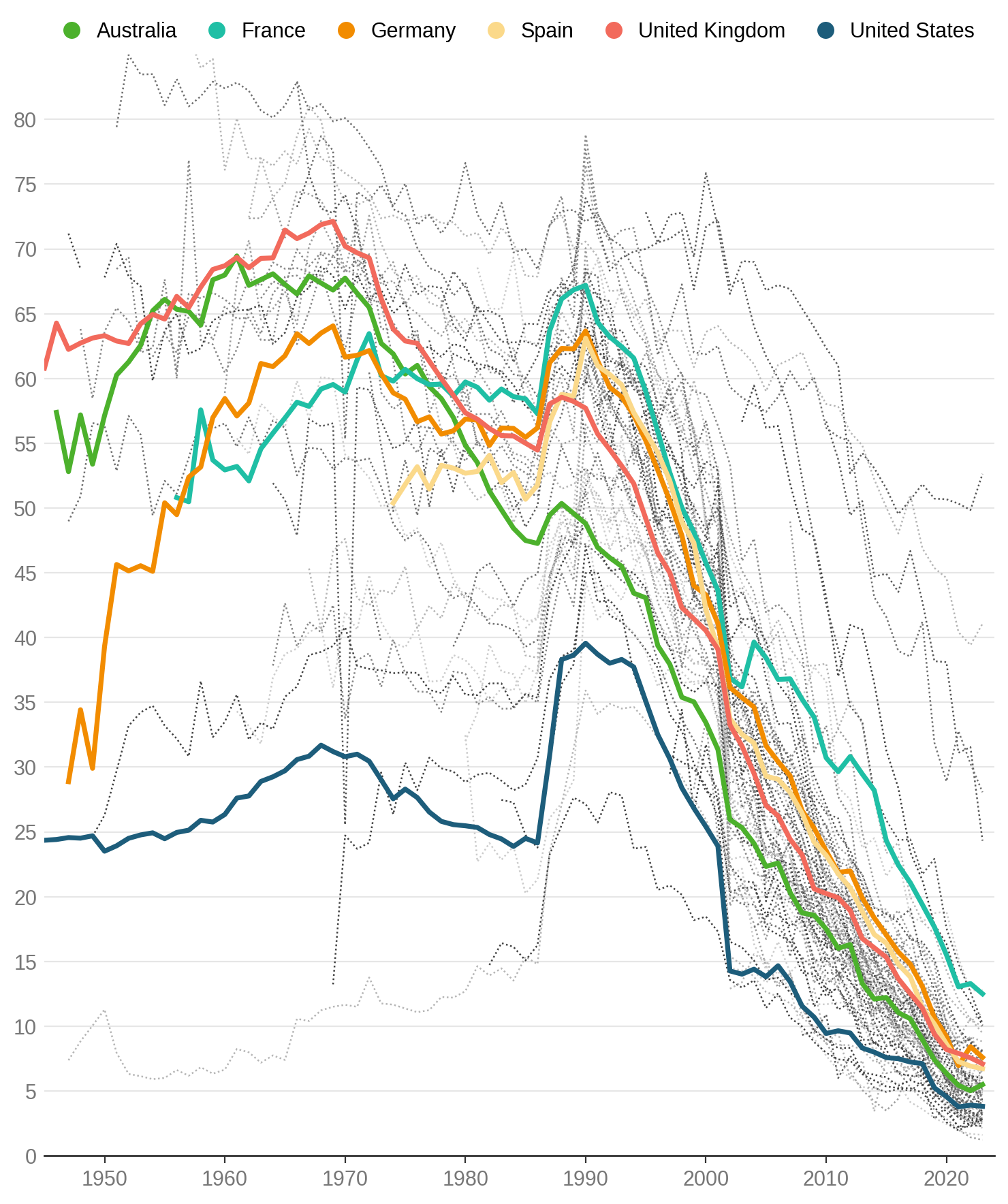}
\caption{Proportion of papers only displaying initial-form names broken down by countries of author affiliations from 1945 until present day.  The grey dotted lines show all countries in the list for context, selected Western countries are picked out in colour. A paper is counted once in every country in which an author's first-listed affiliation lies, and is classified as initial form only when all of its authors use initial form; no fractional weighting is applied.  Countries are drawn only in years with more than 500 papers.}
\label{F9}
\end{figure}

The data in Fig.~\ref{F7} are averaged across all countries but, over the time period that we are examining, not all countries participate in the research enterprise equally.  Thus there is an implicit weighting in the global picture toward specific countries.  If we speak about cultural, political, linguistic or geographic trends then it is important to understand the level of participation of different countries to an international average.  As such, Fig.~\ref{F8} reveals the contribution of each country to global paper production rates in \textit{Dimensions}.  The graph is produced by taking each paper and partitioning it by the countries of the affiliations of each author on the paper and then summing these contributions over all journal articles published in each year.  In the plot, we have only plotted the top 13 countries and have conglomerated under the title "Rest of World" as a 14$^{\rm th}$ participant. As we can see from the Figure, during the period from 1945 to 1990, an extremely high percentage of global research output emanated from the US and the UK.  While Japan and Germany make a significant and sustained contribution from the mid-1950s, between 48\% and 57\% of global output in the 1970s and 1980s came from the US and the UK - English-speaking countries with a somewhat shared cultural base.  Thus, the behaviours summarised in Fig.~\ref{F7} are dominated by an Anglo-American societal behaviours.  However, as we see in Fig.~\ref{F9} cultural alignment between the UK and the US may be more divergent than expected.

Nonetheless, each country has its own social norms and traditions.  Some countries are sufficiently large or diverse that there are different norms within the country while other countries have kinship with neighbours and there is some homogenisation between cultures.  In our analysis here, we see some of these facets emerge naturally from the data.

\begin{figure}[h!]
\centering
\includegraphics[width=.95\linewidth]{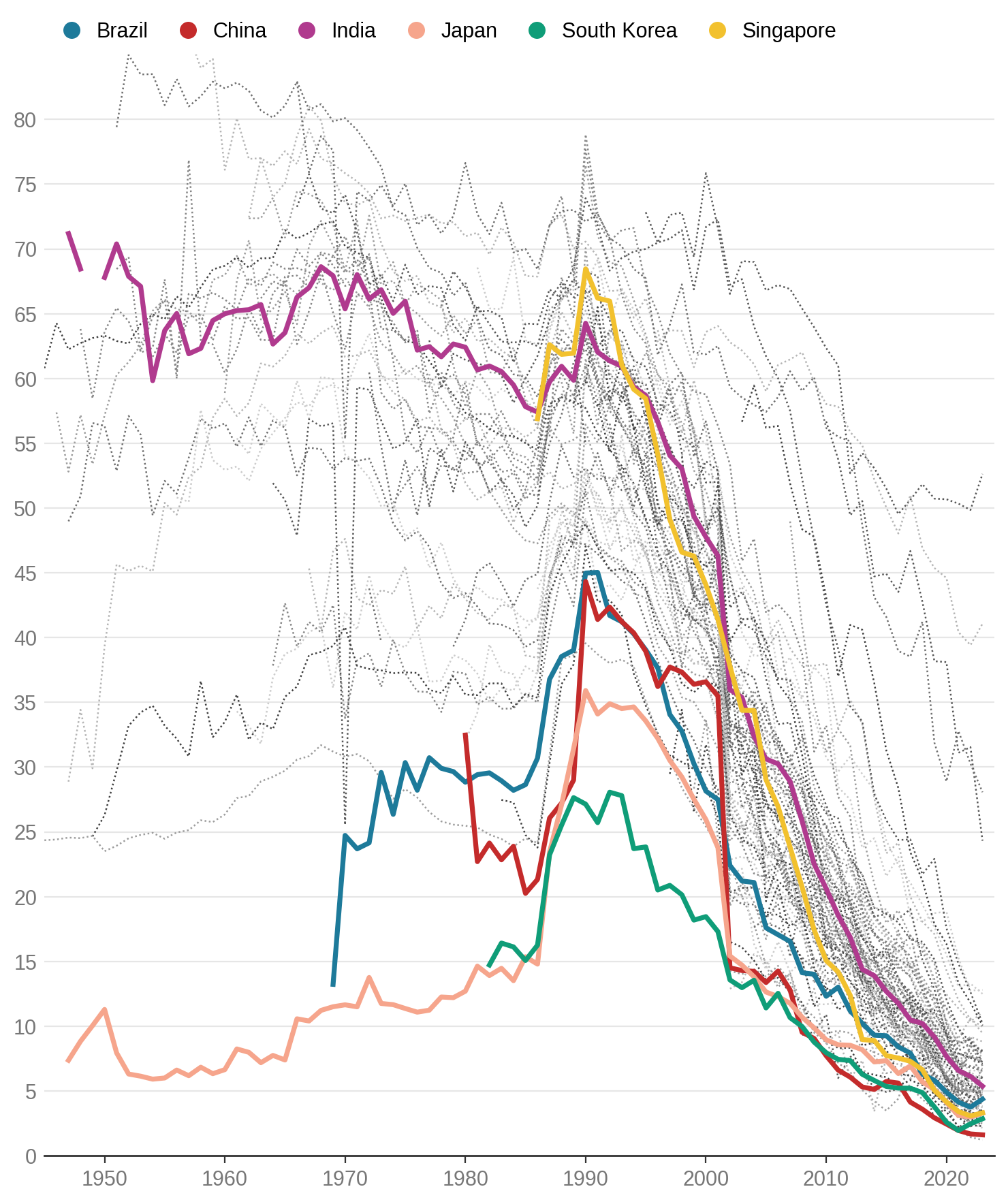}
\caption{As Figure~\ref{F9}, with countries in the Asia-Pacific region and Brazil picked out in colour. The grey dotted lines show all countries that consistently participate in more than 500 papers per year, plotted for context; counting is as in Fig.~\ref{F9}.}
\label{F10}
\end{figure}

\begin{figure}[h!]
\centering
\includegraphics[width=.95\linewidth]{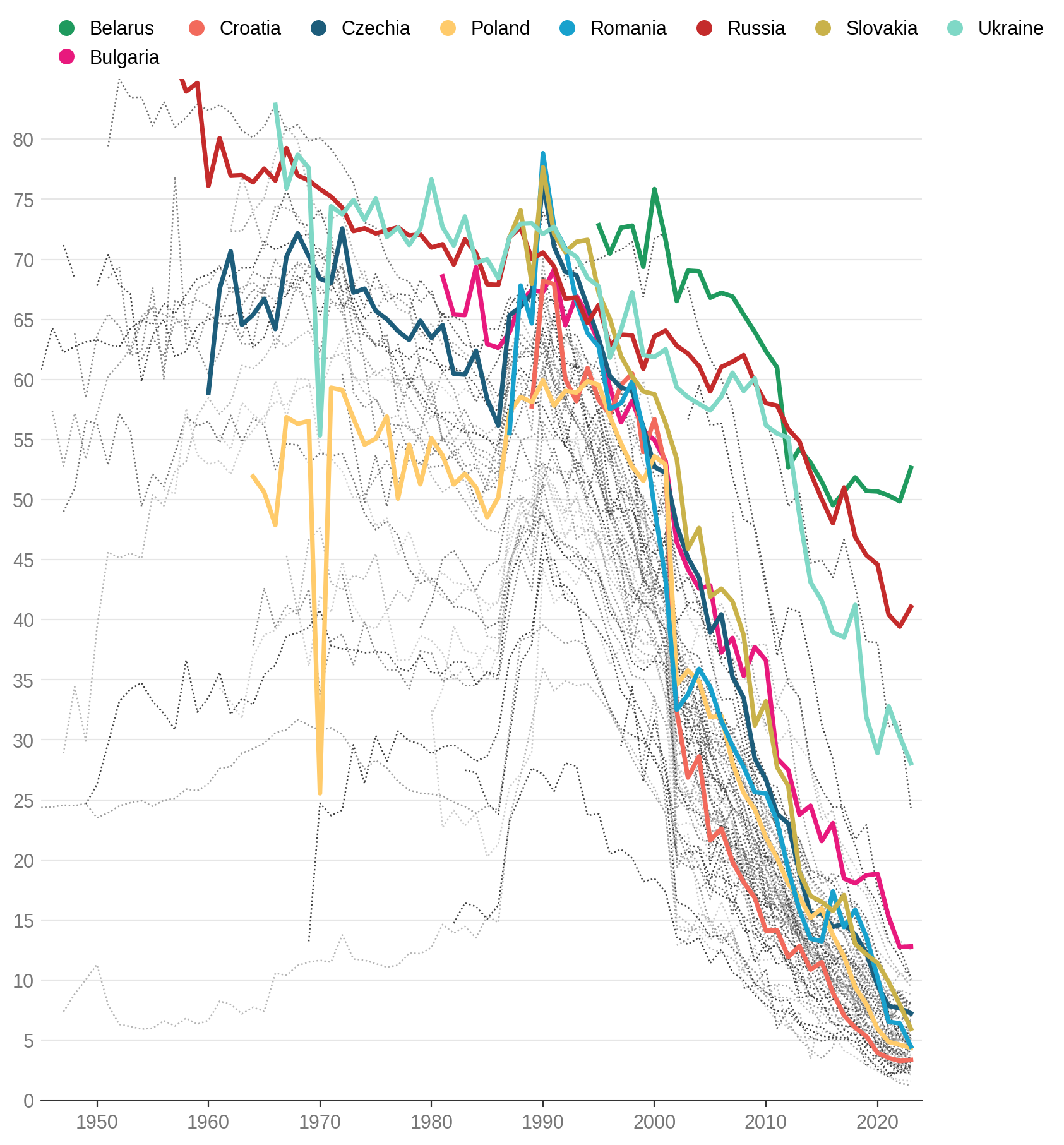}
\caption{As Figures~\ref{F9} and \ref{F10}, with selected Slavonic-language countries picked out in colour. The grey dotted lines show all countries that consistently participate in more than 500 papers per year, plotted for context; counting is as in Fig.~\ref{F9}.}
\label{F11}
\end{figure}

The following three figures (Figs~\ref{F9}, \ref{F10} and \ref{F11}) show the proportion of authors publishing their name using initial form rather than full form---in each case around 25 countries are plotted (countries are selected to appear in the plot when they consistently participate in more than 500 papers per year) in grey with specific subsets picked out in colour.

In Fig.~\ref{F9} Australia, France, Germany, Spain, the United Kingdom and the United States are picked out in colours.  We see that the United States is the country that most consistently leads on use of full form rather than initial form name forms on papers. Interestingly it shows a sharp jump in formality during the late 1980s followed by a slow reversion to its prior behaviour, ended by an almost symmetric sharp drop in the early 2000s.  However, at all points, the US is the ``familiar'' country of the Western world.  Australian authors have been the second most familiar since the late 1970s, having previously been slightly closer to their cousins in the United Kingdom, who having originally led on formality in this group in the 1940s through to the 1970s, started to be much more central in the group.  While the Germans were highly informal in the 1940s, their increase in use of initial form rose significantly in the 1950s and 60s but has since settled close to the middle of this group.  Finally, French authors have been probably more consistent than others in their embrace of the initial form, resulting in their moving from being one of the least formal to the most formal in the group while staying at more or less the same level of initial form usage between 1945 to 1990.  They have then been consistently the highest user of initial form since around 1980 to present day.  Of course, for all members of this group there has been a significant harmonisation over the last 20 years with extremely low levels of initial form usage in place today.

In Fig.~\ref{F10} we have selected countries from the Asia-Pacific region and Brazil.  In this case we see clearly the formalism (initial form alignment) of India from 1945 to present day with it being consistently (and in much of this time period significantly) above its comparators. Brazil passed the production threshold for this analysis in the late 1960s, at which point it was already more aligned to full name usage than most of the rest of the world (below most lines in the grey background), and continues to be even today.  China, Japan and South Korea have all consistently been amongst the countries least likely to use initial form for the longest period of time.  The facet in common for these countries is their use of non-Roman writing systems.  Thus, in native language, author names are often single character, but when translated into English as choice is made to list the name as  as a full name rather than just as a single initial.  This may, in part, be due to cultural and linguistic factors in many Asian countries that give rise to common last names, the lack of a second or third given name, and hence the need to state full first names in order have the ability to disambiguate authors in a pre-ORCID environment \citep{finegan_regional_2004, velden_resolving_2011, liu_study_2012}.  All these issues are compounded by transliteration in the scholarly record.

In Figs~\ref{F9} and \ref{F10}, it is notable that most countries that we reviewed have, at least in recent times, been below the modal behaviour regarding initial form usage.  Thus, in Fig.~\ref{F11} we explore the more formal part of this diagram, picking out Slavonic-language countries.  Not all countries in this group speak a strictly Slavonic language, specifically, Romanian is an Eastern Romance language.  Some countries in the group use the Roman alphabet while others use the Cyrillic alphabet, however, both of these alphabets are structured in a way so as to allow the delineation of first names and initials in a way that some of the Asian writing systems do not.  However, the shared history of these countries (with the exception of Croatia) is that they formed part of the the orbit of the USSR before its fall in 1991.  

A characteristic of these countries was the establishment of strong national academies, which may have led to greater harmonisation of norms of research culture at a national level. What is less clear is whether this is the core effect as this could also be due to pre-existing (i.e. non-research-specific) cultural and linguistic alignments. The data in Fig.~\ref{F11} suggests that these countries have tended to be more formal in their use of initial form rather than full form in research discourse.  Isolating the effect of national academies in research culture agenda-setting is beyond the scope of the current work.  However, we conjecture that this may be a factor in determining behaviour.  In the UK, the diffuse nature of the national academy (being split between the British Academy, the Royal Society and the Royal Academy of Engineering) has tended to mean that universities are more independent and powerful in their own right; in the US a similar diffuse system exists with a much more powerful university sector; and, in Germany the Max Planck Society, the Leibniz Association, Fraunhofer Society, Helmholtz Association and others provide a diverse setting for research culture to develop.  Whether France's more formal preference is due to linguistic structure or administrative structure is, again, hard to disinter---one may, after all, be a consequence of the other.

\begin{figure}[!h]
\centering
\includegraphics[width=.95\linewidth]{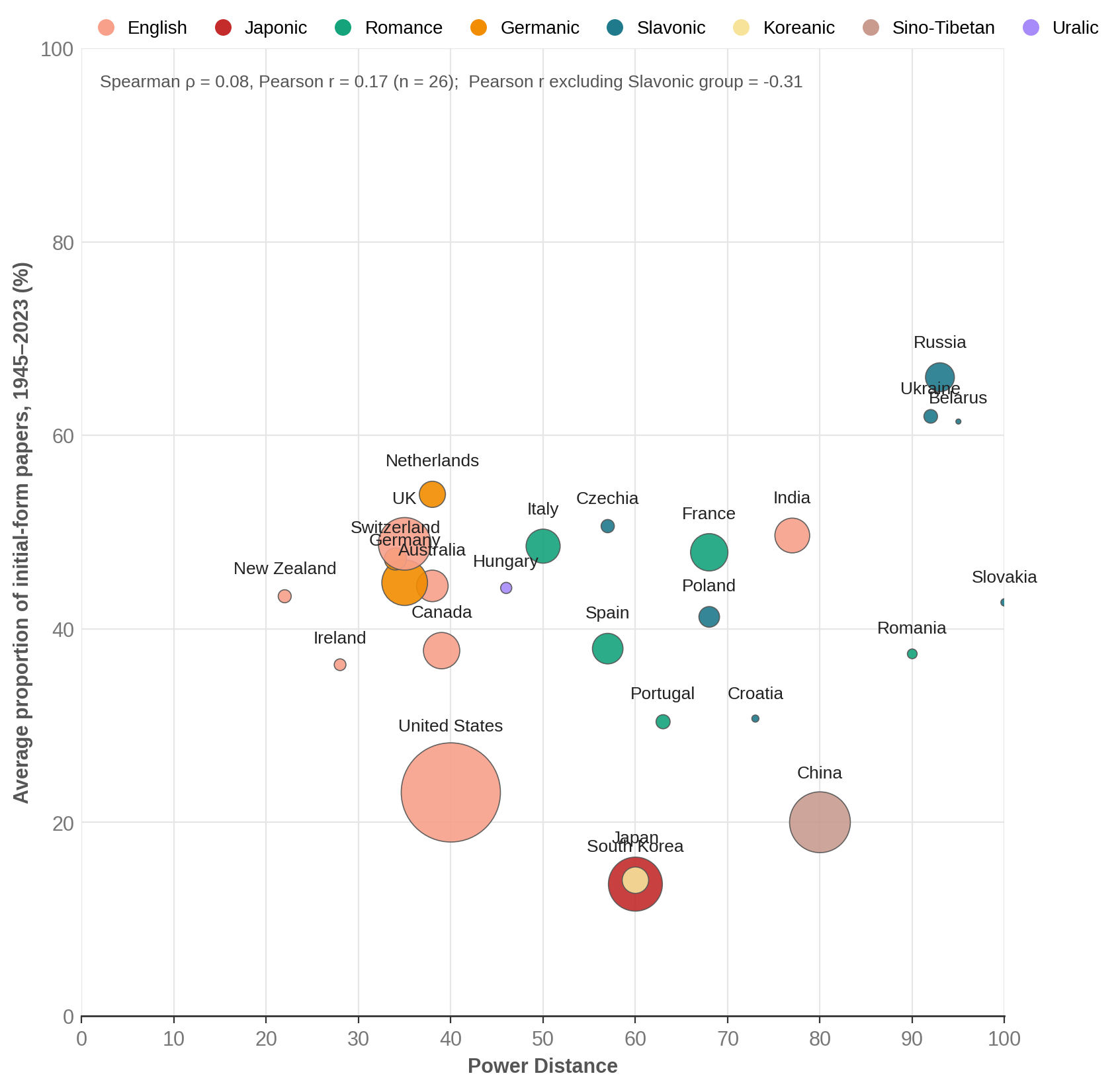}
\caption{Mean, over 1945--2023, of the annual proportion of papers in which all authors use initial form, versus Hofstede's Power Distance Index for a selection of countries (years with fewer than 500 papers are excluded from the mean).  The size of each disk represents the cumulative volume of publications associated with the country between 1945 and 2023.  Each disk is coloured according to the linguistic grouping of the principal language spoken in the country of the author's affiliation.}
\label{F12}
\end{figure}

Overall, over the last 20-year period we see a similar rate of harmonisation toward a more familiar norm.  However, it is interesting to note a few of the journeys that we see in these data. For instance, it is noteworthy that Ukraine's curve mirrors that of Russia throughout the period that we studied until circa 2014.  Likewise, it is interesting to note that Belarus has continued to be consistently formal in its name format.  

Europe is a particularly challenging setting for analysis of this nature due to the richness of the history of the region with significant changes in borders and definitions of countries over the period in which we are interested.  In the Initial Era, some stability prevails but this is set against a backdrop of multinational history in which many countries in Europe have sub-populations of differing language and ethnicities - in some sense countries are an artificial construct.  And yet, they are also the basis of national academies, evaluation systems, and systems of funding.  Yet, at the same time, the Initial Era finds its genesis coincidentally at the time when the Marshall Plan was rebuilding Europe.  A time of austerity for many countries in Europe, when there may have been a propensity toward more formal styles.

Beyond this, countries are also the places of education and the level of formality is instilled through institutions of learning and higher learning.  Another confounding factor in our analysis is that in a more globalised world researchers are more mobile.  Thus, the academic affiliation on their papers may not be indicative of their cultural, linguistic or educational background.   Even more challenging, with the rise of international collaboration, it is unclear whether researchers will naturally import the style of other researchers with whom they have worked, who may not share their cultural background but who may convince them to adopt a different cultural norm.

Our data do seem to suggest some level of cultural, historical, linguistic or political coupling. For example, we see the close mirroring of practices between Ukraine and Russia until 2014, when political developments may have had cultural consequences, on the other hand we see a close relationship between Russia and Belarus seemingly not altering Belarus's more conservative approach to publishing with a preference to initial form.  Indeed, Belarusian-affiliated authors appear to be the sole example that do not follow the overwhelming international norm toward the use of full name on publications.  However, it is very difficult to draw any solid conclusions that are not purely anecdotally-motivated from these data.

In an attempt to gain further insight, we explore the interplay of linguistic and cultural effects further by turning to the work of the cultural theorist Hofstede from the late 1960s. Through his work at IBM, Hofstede developed Cultural Dimension theory \citeyearpar{hofstede_cultural_1983,hofstede_cultures_2003,hofstede_dimensionalizing_2011} in an attempt to understand how cultural considerations affected organisational structures.  One of the cultural dimensions that Hofstede developed, called the Power Distance Index, attempts to quantify the extent to which the less powerful members of organisations and institutions accept and expect power to be distributed unequally.  As such, Power Distance may be thought of as a proxy for how formal a country is culturally in a professional setting.  Of course, such things change with time and the Power Distance shown here is a snapshot at a particular point in time.  We argue, however, that these things do not change as quickly from a cultural perspective as they do from a technological perspective.

Figure~\ref{F12} plots the proportion of authors using an initial form aggregated at a country level against Hofstede's Power Distance for that country.  If there were to be a direct positive correlation between Power Distance and use of the initial form of the name on papers, then we would see country disks clustered along a line rising from left to right.  However, in our plot we see little to no correlation: across the 26 countries the Pearson correlation between Power Distance and initial-form usage is 0.17 and the Spearman rank correlation 0.08, and the Pearson correlation becomes negative ($-0.31$) once the seven Slavonic-language countries are excluded.

Countries with similar languages have similar power distances but there is little correlation with the proportion of authors using initial form.  Indeed, countries with a first language of English have a Power Distance of around 40 but spread from 23\% to 50\% in their use of initial form; on the other hand Romance-language countries are associated with a wide range of power distances (50 for Italy to 90 for Romania) and a similarly wide spread of initial form percentages, from 30\% to 49\%.  However, Belarus, Russia and Ukraine are notably more formal (high power distance) and exhibit higher percentages of initial form usage (61\% to 66\%) than other countries.

This suggests that formality of the culture in which a piece of research is carried out is not a dominant factor in the adoption of initial form as a publication form.  Indeed, since research is typically a highly collaborative endeavour we see people from different nationalities and cultures doing their work in different cultures; furthermore we see collaboration on single papers between different cultures and different locations.  Nor does language or size of research output appear to have a particular effect.  Nonetheless, the curves in Figs~\ref{F9}, \ref{F10}, and \ref{F11}, share similar features and overall shapes in many cases - all the curves trend upward to a more formal style in the mid-1980s, all the curves fall back to less-formal levels after a peak around 1990 and many countries see a precipitous drop in formality in 2000.  This similarity in shape suggests that there is a more fundamental underlying driver than a sociological one.

\subsection{Disciplinary Analysis}
Of course, sociology may differ by subject area more powerfully than by geography.  When we grow up, we are taught certain cultural norms and practices, but the practices that we have for engaging in a research communication context are acquired at university either during undergraduate or postgraduate studies, depending on the field.  In the post-1945 era that we are investigating, university education had already become quite harmonised globally.  The world's ``first-mover'' research nations had controlled and established a set of norms that new entrants have sought to emulate in order to engage on the same footing. (Figure~\ref{F8} demonstrates the extent to which research would, de facto, take place in the English language and in the format chosen by the US, which in and of itself have been heavily influenced by its UK/European ancestry.)  This included adoption of Bacon's scientific method, the concept of the Humbolt institution and the format of scholarly communication in terms of the book and the journal article.  In light of this harmonisation it is less surprising that we see correlations with national attitudes in publication practices and that it is more likely that we should see correlations elsewhere.

\begin{figure}[h!]
\centering
\includegraphics[width=.95\linewidth]{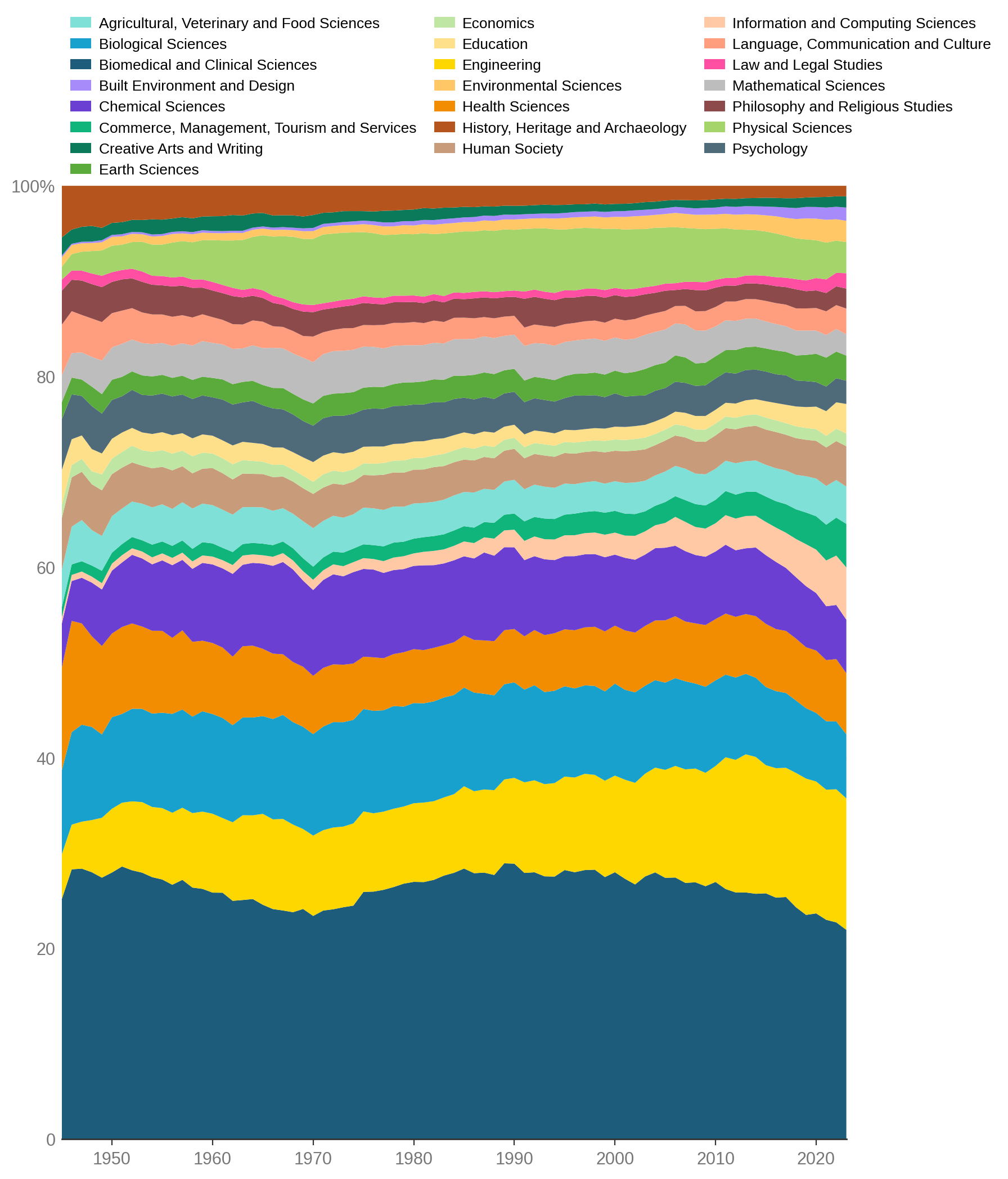}
\caption{Evolution of contribution of different fields to the overall academic corpus with time.  Attributions of papers to fields as per ANZSRC Field of Research Codes from \textit{Dimensions}.}
\label{F13}
\end{figure}

\begin{figure}[h!]
\centering
\includegraphics[width=.95\linewidth]{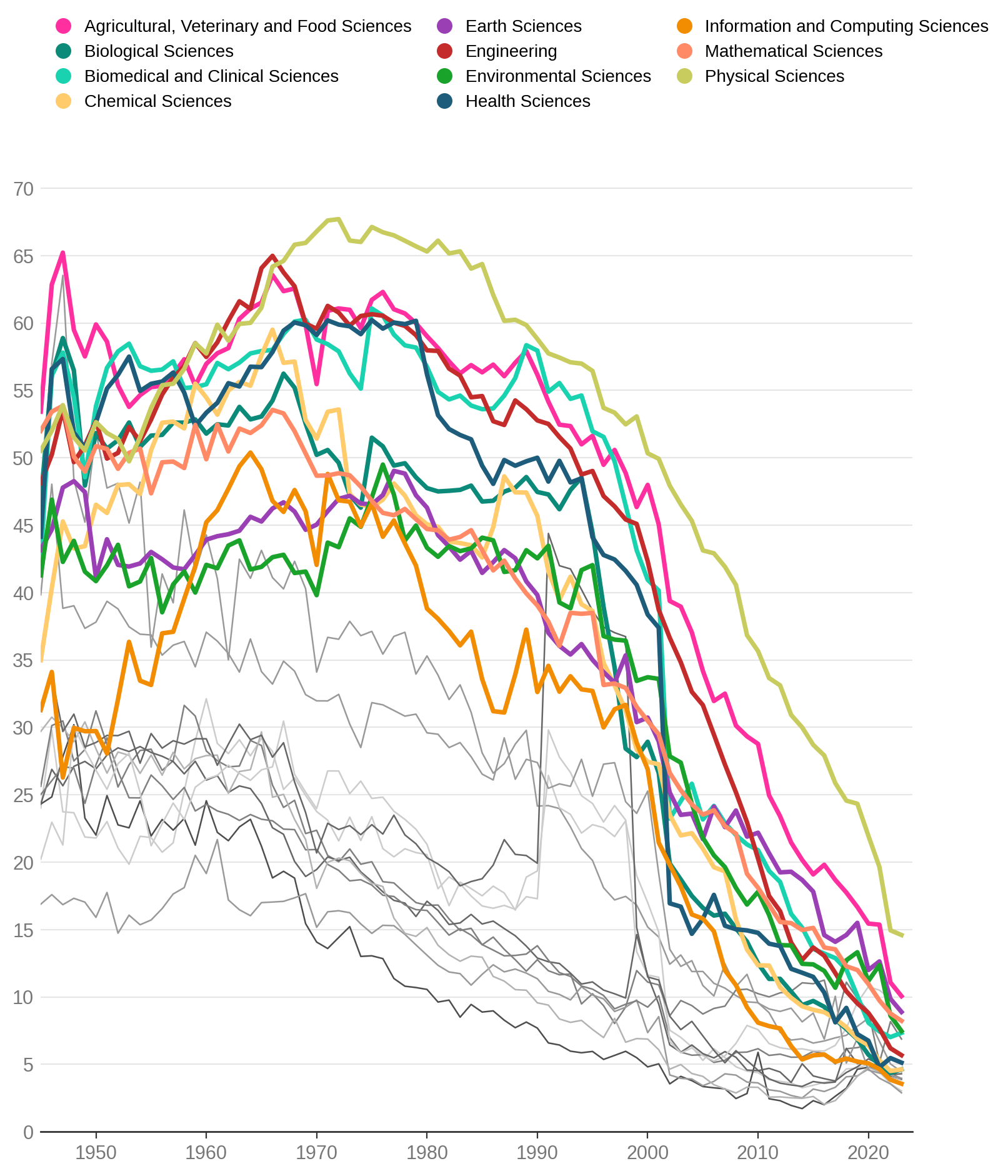}
\caption{Proportion of journal articles in which all authors use initial form, by ANZSRC FoR division and year, 1945 to 2023. Selected Science, Technology, Engineering and Medicine (STEM) subjects are picked out in colour. Fractional apportionment is not applied - there is duplication of counting if papers cross 2-digit-FoR classifications. Fields are determined via \textit{Dimensions} automated attribution to ANZSRC 2020 coding.}
\label{F14}
\end{figure}

\begin{figure}[h!]
\centering
\includegraphics[width=.95\linewidth]{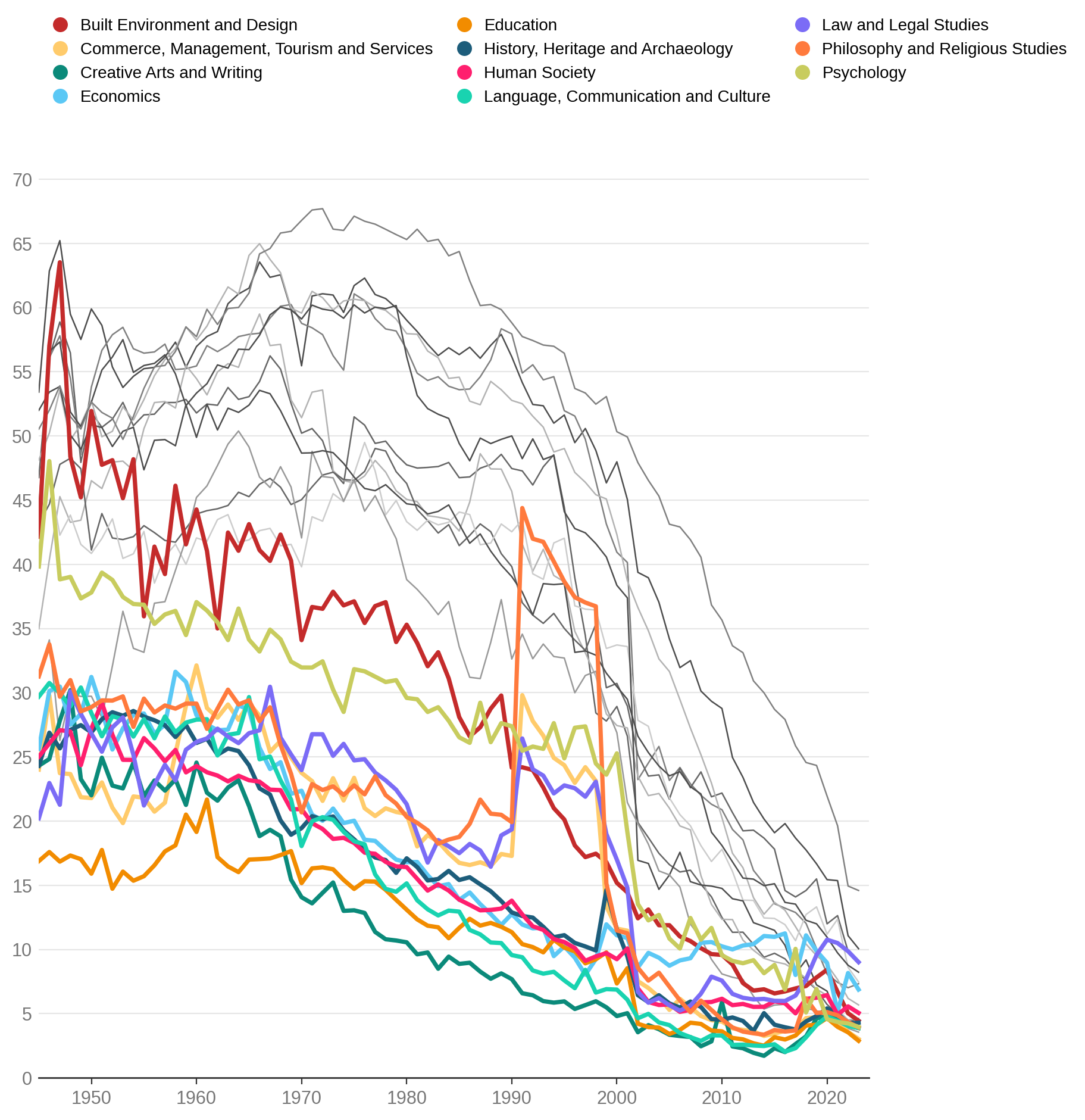}
\caption{Proportion of journal articles in which all authors use initial form, by ANZSRC FoR division and year, 1945 to 2023.  Selected Social Sciences, Humanities and the Arts for People and the Economy (SHAPE) subjects are picked out in colour. Fractional apportionment is not applied - there is duplication of counting if papers cross 2-digit-FoR classifications. Fields are determined via \textit{Dimensions} automated attribution to ANZSRC 2020 FoR coding.}
\label{F15}
\end{figure}

Figures~\ref{F14} and \ref{F15} show analogous plots to Figs~\ref{F9}, \ref{F10}, and \ref{F11} but instead of breaking down the initial form percentage into country contributions, the basis has been changed to examine subject areas.  In this case we make use of the 2020 ANZSRC Field of Research Codes that are automatically assigned to publications in \textit{Dimensions} \citep{porter_recategorising_2023}.  Fig.~\ref{F13} shows the proportion of papers written in each of the subject areas, indicating the contribution level of each subject to the aggregate (or, the effective weighting of each line in each plot).

Figure~\ref{F14}, the first of the subject-based figures, highlights STEM subjects.  It is instantly noteworthy that these subjects cluster to the top of the graph, indicating a greater use of the initial form in the publications.  The physical sciences appear to have the highest adoption of this format with information and computing sciences have the lowest level---both areas consistently hold these positions. Again, all fields have a similar overall shape, peaking in the late 1960s and 1970s, since which the use of the initial form standard has declined.  Interestingly, while the precipitous drop in 2000 still exists for some subjects the way that it is in the country data, a variety of fields do not show this feature, including \textit{Physical Sciences}, \textit{Agricultural, Veterinary and Food Sciences}, and \textit{Engineering}.  

The second subject-based figure focuses on the SHAPE disciplines.  With the exception of Built Environment and Design in the 1940s and 1950s and, in a brief period from 1991 to 1998, Philosophy and Religious Studies, all these fields exhibit lower levels of use of the initial form. As with the STEM-subject view, we see the 2000 precipitous change in the use of the initial form format is only significantly visible in Philosophy and Religious Studies, with Psychology showing this change to a lesser extent.

In C. P. Snow's Tale of Two Cultures \citep{snow_two_2012}, he argues that Social Sciences, Arts and Humanities (which we will refer to as the SHAPE disciplines) have a different culture to Science, Technology, Engineering and Medicine (which we will refer to as the STEM disciplines), and this appears to be born out in our data.

With the exception of \textit{Built Environment and Design} before 1960 and \textit{Philosophy and Religious Studies} for a brief period in the 1990s, the SHAPE disciplines lie below the STEM disciplines consistently in their use of initial form.  When we take account of the weighting of volumes in different disciplines shown in Fig.~\ref{F13} then it is clear that the average of disciplines will be weighted toward the disciplines of \textit{Biomedical and Clinical Sciences} and \textit{Health Sciences} and \textit{Biological Sciences}.  All three areas are aligned with medicine and hold a relationship to the formality associated with healthcare professionalism.  At the same time they are three of the largest disciplines by number of outputs, accounting for around 45\% of global publications for most of the period, falling to 35\% by 2023. 

\subsection{Journal analysis}
\label{sec:journal}
It is of course impossible to consider the evolution of articles, without also considering the evolution of practices at the journal level.  Journals have historically been associated with scholarly societies that are disciplinary in nature and have embodied communities and have been reflections of their practice. Not only this, journals are the wrapper for article publications with journal management teams having responsibility for technological choices such as how the move from print to online was accomplished and when it took place. In the digital era, they have also held the choice of specific platforms and hence have had their options defined for them by suppliers of these technologies. Not limited to the digital era of publishing, journal editors have made choices such as the adoption of policies such as house styles which opine on the use of grammar, and of author name styling. Understanding these effects in situ is a key input to further enhance our picture of the landscape.

\begin{figure}[h!]
\centering
\includegraphics[width=.95\linewidth]{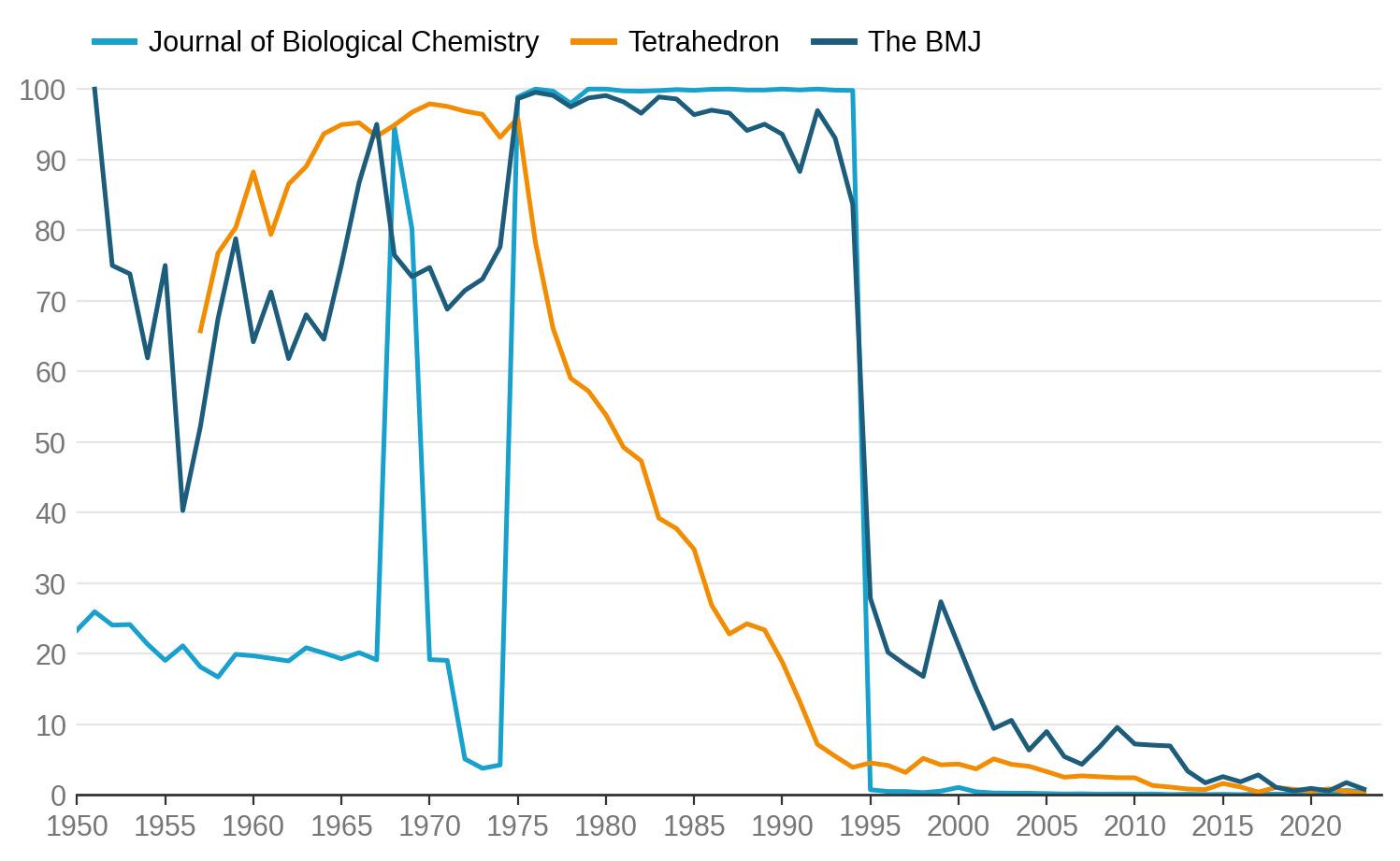}
\caption{Proportion of initial form usage in three journals, Journal of Biological Chemistry, Tetrahedron and The BMJ, from 1950 to 2023: for each year, the mean over research articles of the share of a paper's authors in initial form.}
\label{F16}
\end{figure}

Figure~\ref{F16} shows the changes that we see in Fig.~\ref{F14} at the micro level. The examples chosen in Fig.~\ref{F16} are well documented and hence we can see some of the changes that took place through the Initial Era with more of the detail.  

In the mid nineties, Highwire press emerged as a significant player in the online hosting of Journal content. Starting with the Journal of Biological Chemistry, Highwire was at the forefront of re-imagining the definition of a paper from a digital `photocopy' of a physical paper, Highwire's technology allowed the paper to be a fully interactive digital object in a contemporary sense---text was textual rather than graphical and hence was searchable, elements of the body of the document were broken down into component parts so that images and their captions were distinct objects in the paper \citep{noauthor_highwire_2000,watson_it_2011}. 

By 2000, the Journal of Biological Chemistry noted that, following the lead of Highwire press, many journals now considered the online version of the publication to be the version of record.[2] As Fig.~\ref{F16} shows, the precipitous switch from initial-form to full-form names in 1995 coincides precisely with the launch of the journal online, suggests that the new system made first names mandatory. 

The BMJ makes a similar shift, also in 1995, although its profile suggests that that first names are strongly preferred but not enforced \citep{smith_preparing_1996}. Prior to this, in around 1975, it appears that all three journals, updated their editorial policies as there are sharp changes of behaviour. Both the BMJ and the Journal of Biological Chemistry appear to have put journal policies in place that mandate (or close to mandate) initial form, while Tetrahedron appears to have either removed a mandate for initial form or even actively promoted full name form given the initial rapid drop off from 96\% to 54\% initial-form name usage within just five years from 1975-1980.

By 2016, the BMJ had switched to using ScholarOne for manuscript submission. Within this workflow, author names are not keyed in individually, but, like a CRM system, selected from a database of previously recorded identities. If a person cannot be found (by email address), a new record can be created, and given names and last names are required. The system is normative in the sense that no indication is given that a given name could just be an initial. Authors are not so much recorded against the article as people are linked. A single representation of an author is now used across all journals that use the ScholarOne service. The BMJ's usage of initial-form names falls to around 2\% from 2014 and to around 1\% from 2018.

The effect of technology as a driver of change is clearly delineated in this plot as Tetrahedron declines gradually indicating a change in social preference or weaker forms of change such as journal policy and house style changes. Tetrahedron did not go live with its first online submission system until 1995, so it is assumed that this change was managed offline by typesetters.

\subsection{Technological Analysis}
\label{sec:tech}
We may reasonably ask what the impetus was for the change in behaviour in the medicine-related fields (the significant, sharp discontinuity that we see for Biomedical and Clinical Sciences, and  Health Sciences in Fig.~\ref{F14}). The speed of change suggests that this was a technological rather than a cultural driver, since technological changes tend to be implemented with greater speed.  Indeed, since the medical fields share infrastructures such as MedLine and PubMed, which are funder orientated, there could be a strong funding-aligned impetus that we can pinpoint in this case. 

In order to explore PubMed we need to find differentiating characteristics of the PubMed data to be able to track effects. Helpfully, PubMed and its anticedents such as MedLine have conferred unique identifiers to articles prior to the advent the generalised use of Crossref DOIs. This means that some articles on PubMed do not have DOIs and only have a PubMed identifier.

\begin{figure}[h!]
\centering
\includegraphics[width=.95\linewidth]{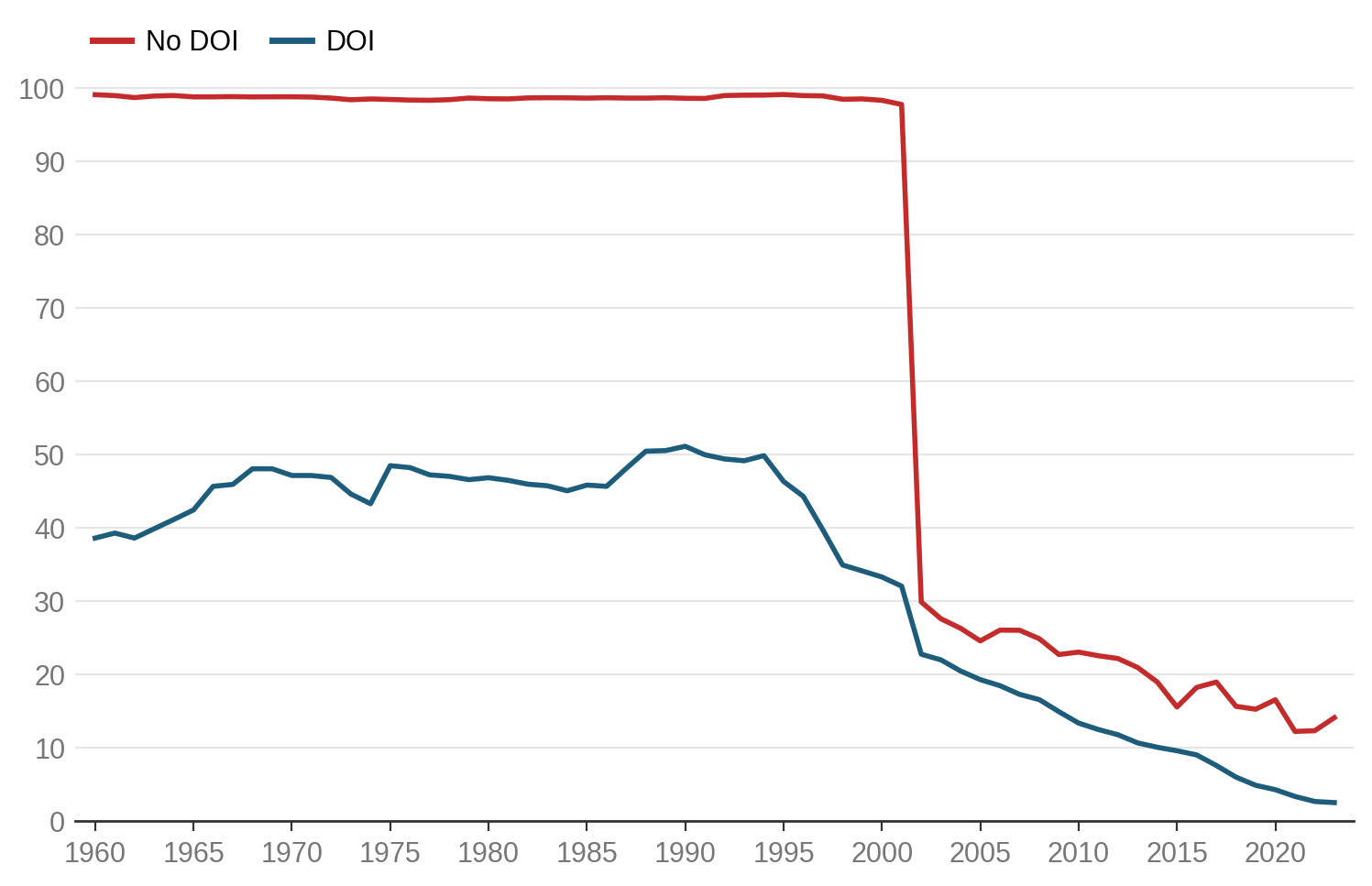}
\caption{Percentage of author positions in initial form on PubMed-indexed articles with (blue) and without (red) a DOI, 1960 to 2023.}
\label{F17}
\end{figure}

Figure~\ref{F17} shows distinct and different behaviours between authors associated with articles appearing on PubMed that are associated with DOIs and those which aren't.  The blue line shows the percentage of authors associated with a DOI where initial form is used; the red line shows the same percentage for authors associated with articles without a DOI.  For articles with a DOI, initial-form usage rises to around 50\% by 1990 and then changes gradually to lower rates as authors migrate to full form.  We note a marked drop in the blue line in 2002, from 32\% to 23\%, which, we speculate, would be associated with the launch of PubMed's new platform.  This new platform supported given name metadata fields and hence naturally encouraged the use of full form names \citep{nlm_medline_2002,nlm_skill_2009}. 

Articles without a DOI (only having a PubMed identifier), shown in the red line of Fig.~\ref{F17}, have a distinctly different shape.  The difference in behaviour of the red line is explained by a systemic development: When Crossref introduced the opportunity to implement DOIs for academic articles, their metadata format was sufficiently fully formed and modern to allow for a field to include given name and publishers adopting the new standard were able to update their back catalogues with the additional data.  Thus, the blue line is a representation of the actual metadata included on articles as they were published, whereas the red line is the result of the legacy metadata standard that existed prior to the advent of Crossref's metadata schema \citep{crossreff2009}.  Even though articles without a DOI may, in fact, have full form names on the publication this is not represented in the metadata.

Similar parallels in behaviour can be seen with other major bibliographic and bibliometric databases that existed through this period, with Web of Science collecting first names from records processed after 2007, and allowing them to be searched in 2011 \citep{clarivate_web_2022}, which Scopus had done in 2004 \citep{hane_elsevier_2004, sullo_scopus_2007}. 

The ongoing decline in the use of initial form from 2002 to present day, may not be due only to technology changes but rather may be motivated by changes to another scholarly institution---research evaluation.  In the period from 2002, technological developments took place to introduce of (current research information systems) CRISes \citep{jorg_cerif_2010}, to address the need for reporting data in research economies that introduced research evaluation such as the UK and Australia.  ORCID was incorporated in August 2010 to tackle the disambiguation problem at a deeper systemic level, though the registry itself did not launch until October 2012 \citep{orcid_first_decade_2022} - perhaps surprisingly the registry's launch does not seem to have a particularly direct effect on these data \citep{butler_scientists_2012,schiermeier_research_2015,meadows_persistent_2019}.

\subsection{Gender Analysis}
Before presenting the gender-focused analysis, we should be clear about what it can and cannot show. By construction, gender attribution from author names is only possible for authors whose papers display full given names; authors who publish in initial form are gender-indeterminate to any name-based method. The analysis that follows is therefore \textit{not} a measurement of gender participation in research, but an analysis of \textit{gender visibility in the scholarly record}---of when, and to what extent, the metadata permits gender to be inferred at all. We also acknowledge the well-documented limitations of name-to-gender attribution methods more generally; Lockhart et al.~\citeyearpar{lockhart_name-based_2023} provide a critical treatment of these issues. Our use of genderize.io is not robust at the level of any individual author, and the analysis here makes no claims about specific researchers.

Finally, we turn our attention towards a gender-focused analyses.  We have pointed out reasons that gender may be hidden by choice in certain contexts in the introduction to this paper. However, it is clear that the reasons for  the obscuring of gender in research publication are considerably more subtle \citep{lopez_lloreda_women_2022, teich_citation_2022}.  Countless studies demonstrate differences in outcome for women in peer review in both grant and publication contexts, and citations of their work is lower than for male counterparts \citep{zhou_systematic_2018,draux_gender_2018, draux_gender_2019,holmes_gender_2019}. Further work seeks to understand the causal nature of these effects \citep{traag2022causal}.  What we show here is that our data analysis as a further dimension for consideration namely that, perhaps unsurprisingly, technology can be a significant modifier to gender visibility in publication.

\begin{figure}[h!]
\centering
\includegraphics[width=.95\linewidth]{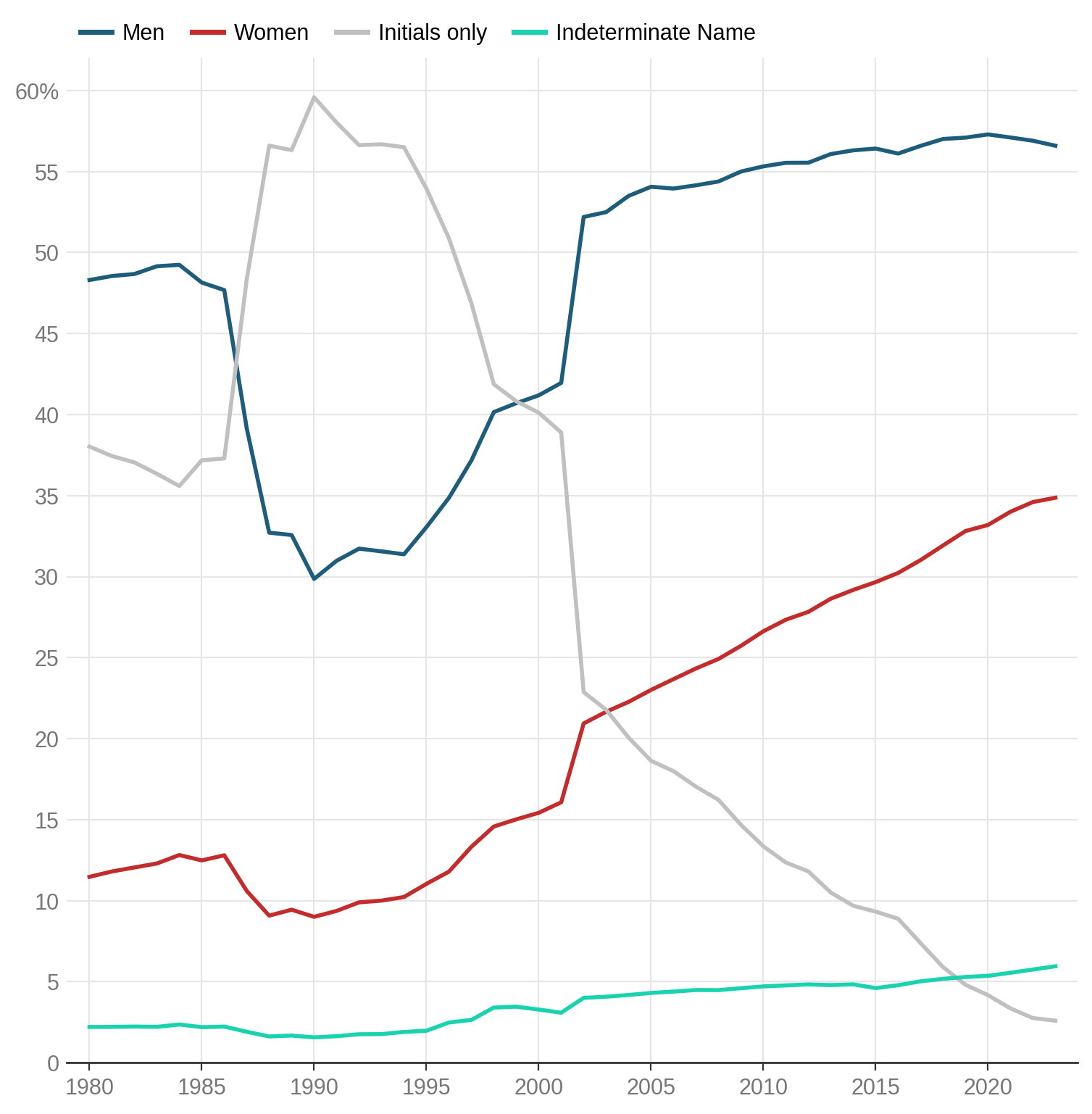}
\caption{The development of perceived gender of authors on PubMed-indexed research articles from 1980 to 2023, as the mean over papers of the share of each paper's author positions in each category (the four categories sum to 100\%).  Blue line: names statistically associated with men; Red line: names statistically associated with women; Grey line: initials only; Green line: a given name that genderize.io does not associate with either.  Gender attribution uses genderize.io with the author's first-listed affiliation country.}
\label{F18}
\end{figure}

Figure~\ref{F18} illustrates the development of the perception of gender in authorship from 1980 to 2023 in the context of PubMed.  As we have seen in Fig.~\ref{F17} there is a distinct technological event in 2002 that caused a shift away from initial form and toward full form names.  We see that both the blue line (names statistically associated with men) and the red line (names statistically associated with women) jump upward around 2002, at which point there is a steep decline in initial form usage (grey line).  Increased representation of full names has given rise to a modest increase in indeterminacy (green line), from around 3\% to around 6\% of output.  The rise in initial-form authorships between 1987 and 1995 (grey line) coincides with a sharp growth in the number of PubMed-indexed research articles carrying affiliation data in \textit{Dimensions}, from around 130,000 per year in 1986 to 290,000 in 1995, and with the near-universal use of initials on records without a DOI (Fig.~\ref{F17}); we read it as a change in the composition of the indexed record rather than in authors' practice.

While the step in the blue line in 2002 draws the eye it is important to take this change as a proportion of the population---a move from 42.0\% of authorships with names statistically associated with men in 2001 to 52.2\% in 2002 is a 24\% increase.  However, for names statistically associated with women, the move is from 16.1\% to 20.9\% - a 30\% increase.  In addition, in the 20 years from 2002 to 2022 names statistically associated with men move only from 52\% to 57\%, whereas names statistically associated with women move from 21\% to 35\%.

While the reporting of gender participation in research is still relatively nascent with systematised aggregation of reporting only available since around 1996. Already in many regions of the world that reported in 2002, the date of the change in PubMed systems, women regularly accounted for between 30\% and 50\% of researchers \citep{UNESCO2021}. Yet in 2001, the last year before the change, PubMed authorships attribute only 16\% of output to those with names statistically associated with women, while 42\% of authorships were in initial form or otherwise of indeterminate gender.

The 2002 platform change in PubMed therefore functions as something close to a natural experiment: the same population of researchers is being measured before and after a technological discontinuity, with no plausible mechanism for an underlying participation shift in a single year. The disproportionately steeper increase in female-coded names following the change---combined with the fact that the pre-change indeterminacy was substantially larger than the post-change indeterminacy---is consistent with the interpretation that the initial-form convention disproportionately suppressed the visibility of women researchers in the metadata. This is, we emphasise, a finding about \textit{what the metadata could show}, not about \textit{who was doing the research}. But it is a finding with material consequences: a scholarly record that systematically under-represents women's authorship in its searchable metadata is one in which women's contributions are harder to find, harder to count, and harder to credit.  

\section{Discussion}
We have argued that the period from 1945 until 1983 defined a period that we call the ``Initial Era''.  Rather than being a normal period, we believe that it is exceptional in the scholarly record, being the only era that we can identify in which there is an extended period where initial form was used instead of full form names on academic manuscripts.

Each of the subsections in the Results section of this paper examines the data from a different aspect to allow us to, firstly, assess the robustness of the data, and secondly, to draw together a picture of the causes and effects of the Initial Era.  In this section we will use our results to speculate on the likely genesis of the Initial Era and to examine the effects.  We will finish by considering the future of the scholarly record and the potential for future bibliometric archaeologists to perform similar studies.

\subsection{The rise of the Initial Era}
We speculate that the adoption of initial form during this exceptional nearly-four-decade period is far from being accidental.  The entry to this period can be marked as the beginning of what might be thought of as the Vannevar Bush-era \citep{bush_endless_1945} and was a time when science was being established as ``serious business'' with a more formal cultural norm associated with it.  This may have been influenced by the rise in importance of science and medicine and the view that technology could solve all problems.  Similarly, the Marshall Plan rebuilding of Europe could have led to the import of norms the US, but this seems less likely since the US is continually one of the lower users of initial-form (see Fig.~\ref{F9}).

However, during this period, research output was dominated by the US, the UK and Japan---all countries that were predisposed to publish in English and to reflect the cultural norms in the US for political reasons.  The Bretton-Woods Era came to an end in the late 1970s and the Cold War in the late 1980s, and with the rise of globalisation greater international collaboration has been possible, leading to Adam's Fourth Age \citep{adams_fourth_2013}.  Yet, as we see from Figs.~\ref{F9}, \ref{F10} and \ref{F11}, the use of initial form persisted (and even increased, possibly due to the constraints of paper media and their interaction with the era of Big Science) from 1980 to 2000. As demonstrated in our technological analysis, it was a combination of house style decisions, technology changes in the form of PubMed and Crossref, and the switch to online journal submission systems that fundamentally changed research culture back to full name usage. It remains a ``chicken and egg'' issue as to whether technology evolved to meet cultural needs or whether technological change gave cultural norms an opportunity to assert themselves.

The methodology behind our analysis in Sec.~\ref{sec:geo} is prone to reflecting international trends.  By construction, the unit of analysis is the paper and is classified as an initial-form paper only in the situation that all authors use that form.  There are normative trends implicit in a global research community---publishers have house styles that are applied regardless of origin; if the authors who brought funding for the work, or who are in some other way senior, choose a style of publishing their name on the paper perhaps there is a psychological bias that goes on that we cannot track in the data; if a majority of authors choose a specific style then how many authors are comfortable breaking step and choosing a different form for their name?  International collaboration is another anchor to which our methodology is particularly sensitive as only papers that are entirely initial-based factor in analysis such as Sec.~\ref{sec:geo}---hence, a core assumption of the analysis is that the kinds of effects discussed here were sufficiently strong as to normalise behaviour across a paper regardless of geographic background.

As collaboration has become more of a norm in research (see Fig.~\ref{F3}) and as many countries have invested in and developed their own advanced research economies (see Fig.~\ref{F13}), a much more diverse community has emerged. Hence, once ``senior partners'' are now just simple partners.  On the one hand, this will necessarily require the representation of newly evolved norms.  Yet, the norms are already set and new norms take time to evolve. 

\subsection{The fall of the Initial Era}
As with any era in history, a critical understanding the outcomes and impacts of the period are critical in understanding ourselves and improving current and future practices.  

The fall of the Initial Era invites several plausible accounts. The most parsimonious is technological---that the timing of the decline is closely tied to specific changes in the bibliographic infrastructure (Section~\ref{sec:tech})---but technology alone is unlikely to be a complete explanation. We posit that the trend is best understood as the interaction of three major factors:
\begin{enumerate}
    \item Technological harmonisation: Allowing the choice of capturing and using full names across many different types of system from research information management within institutions to evaluation, and from publishing to grant application and administration;
    \item Globalisation of research: The natural harmonisation of norms as more and more researchers collaborate outside a narrow context;
    \item Trust: The ongoing need to propagate trust in the sense of understanding the identity of participants in the research ecosystem.
\end{enumerate} 

All three of these points play into the greater narrative of how research is changing around us. There is a greater need for transparency as research has an increasingly significant effect on everyday lives and as governments investment more in research, it is critical to make knowledge common. The move to greater transparency is multifaceted---making the knowledge itself more transparently available is just one stream; making its production and provenance transparent is another.  Without transparency of production and provenance, we cannot expect the trust, which is needed for research to progress.

The articulation of this need for openness is often found in evaluation mechanisms, either national or funder based.  All the factors that we identify above relate to evaluation.  Technological developments, especially around name disambiguation, relate to the rise of ORCID as a standard \citep{haak_using_2018,porter_measuring_2022}, the use research information management systems and research profiling tools \citep{ilik_openvivo_2018}.  The globalisation of research has led to increased researcher movement and for researchers to be able to move freely, they need to be able to assert their credentials---which involves asserting identity in relation to their research work. And, in a world where nefarious actors are on the rise \citep{porter_identifying_2024}, understanding the provenance of research often relies on understanding identity.

While one may think that the study of the use of names in academic work is perhaps facile, our study shows that it is anything but trivial. Ultimately, we believe that it is possible to justify the argument that the effect of a period in which the usage of given names was suppressed---the Initial Era---has been to render the contribution of women to research systematically less visible in the scholarly metadata, suppress data about social norms, and increase the vulnerability of the research system to abuse by weakening provenance.

\subsection{Alternative drivers and their interaction with technology}
Beyond the three factors discussed above, several alternative drivers warrant consideration. Editorial policy and house style---the layer at which most authors actually encounter naming conventions---clearly played a role: the \textit{Tetrahedron}, BMJ, and \textit{Journal of Biological Chemistry} trajectories in Section~\ref{sec:journal} show distinct policy events in the mid-1970s and again around 1995 that cannot be reduced to platform-level technological change. Academic prestige economics---the increasing importance of credit, evaluation, and the disambiguation of contribution to those processes \citep{pardo-guerra_quantified_2022}---also pushes toward full-name representation, since initials are a poor basis for the bibliometric and reputational accounting that has become integral to academic careers.\footnote{For a critical view of metric-driven evaluation, see, e.g., Wilsdon et al., \textit{The Metric Tide} (HEFCE, 2015).} Globalisation provides a third alternative explanation: as collaboration networks expand beyond a narrow Anglo-American base \citep{adams_fourth_2013}, the prior conventions of small, mutually-known communities give way to the requirement that authorship be legible across cultural and linguistic boundaries.

These accounts are not in competition with the technological account so much as in conversation with it. The technological changes we document in Section~\ref{sec:tech} created the \textit{opportunity} for full-form names to become normative; the editorial, evaluative, and globalisation pressures supplied the motivation. The Initial Era ended because a set of mutually reinforcing pressures aligned, and our data are best read as evidence that this alignment was distributed across multiple layers of the research system rather than concentrated at any single one.

\subsection{The future of bibliometric archaeology}

If we have learned anything in the writing of this paper, we must reflect that the lens through which we look is as important as the analysis itself. While \textit{Dimensions} is a useful tool in understanding demographics and evolution of certain types of literature, it is a product of its world and of its construction \citep{hook_dimensions_2018}--specifically the requirement of a DOI or other mainstream identifier limits Dimensions' coverage at this time. Data coverage pre-1800 shows low volumes of output and, in part, this is because output levels are lower but it is also because strategic choices have had to be made by infrastructure organisations in what should be afforded, the cost and effort of creating digital versions and digital identifiers for past material that is not likely to be of current research value in a non-sociohistorical setting. This mirrors the more hidden lack of data that we have identified in the Initial Era, where the contribution of women is hidden through the technical implementation of a social construction.

In the current contemporary setting, capturing and retaining information that describes an individual is changing: GDPR regulations in Europe and their equivalents around the world are designed to allow people to be forgotten if they wish.  Yet understanding the identity of those who carried out a piece of research seems natural to many of us and even essential information in order to understand biases that may be present.  If one is thinking purely scientifically about record keeping, then it makes sense to gather as much data as possible.  However, this shows a profound lack of subtlety and even respect when considering issues of identity and ways of knowing---cultural approaches that come from a different tradition than that of Western research but which, as research diversifies its community, need to be included.

It is apposite to ask whether we live in a special time in bibliometric analysis where name data continues to be freely available and whether this period will persist.  ORCID now assigns unique identifiers to any researcher who would like to have one.  We live in a time of increasing polarisation and researchers who wish to carry out certain types of research or express particular opinions can face censure or removal of funding.  If we take the logic of using research identifiers further, some might argue that the adoption of cryptographic identities, like Satoshi Nakamoto, would allow for researchers to regain intellectual freedom.  Zero-knowledge proofs could be used to authenticate claims for career progression, double-blind peer review would be built in to such a system, potentially increasing the overall fairness of both publication and funding practices. Others might argue that this undermines deeply the trust that we have in the scholarly record and permits abusive expression or propagation of fake research or support of political or commercial interests from ``behind a mask''.  This is, perhaps, the modern-day equivalent of the binaries pointed out by Evans et al.~\citeyearpar{evans_craft_2023} in their commentary on the role of the algorithm.

From a purely analytical standpoint, it would be sad to see the scholarly record lose an aspect that is reflective of current trends.  But, perhaps one role for blockchain-like technologies may be in the detailed capture of demographic information of researchers in a way that protects it for future generations in much the same way that detailed census results are not revealed at the time but which are held in trust for a hundred years. While statistical information on gender, culture and background might be helpful in workforce planning and in helping to ensure that the research system carries on working toward greater levels of inclusion and representation, detailed information may be preserved for future historians and bibliometric archaeologists.

More positively, it could be argued that the transformation away from name formalism should not stop at author bylines. Name formalism is also embraced in reference formats. It could be argued that even within a paper, this formalism suppresses the diversity signal in the research that we encounter. Reference styles were defined in a different era with physical space constraints. Is it time to reconsider these conventions - establishing new paths of analysis in the bibliographic record?

Within contribution statements that use the Credit Ontology \citep{allen_brand_2014}, initials are also commonly employed to refer to authors although this is not part of the standard. This convention also creates disambiguation issues when two authors share the same surname and first initials. Here too, as the Digital Structure of a paper continues to evolve, we should be careful not to unquestioningly embed the naming conventions of a different era into our evolving metadata standards. 

\section{Acknowledgements}
The authors wish to thank Briony Fane for her careful reading and commenting on the manuscript, and to Kathryn Weber Boer for her suggestions to improve discussions on issues of gender.

A preprint of this article was posted to arXiv \cite{porterhook_2024} prior to open peer review on the MetaROR platform and subsequent submission to Scientometrics.

\section{Data Availability}
The data and code for this paper are available from Figshare at: \url{https://doi.org/10.6084/m9.figshare.25664154}.  The deposit comprises the BigQuery SQL that builds the analysis tables and each figure's data, the plotting script, the exported data for every figure, and a README giving the mapping from each figure to its query, data file and plotted columns.  The data used to create this version of the paper were taken from \textit{Dimensions} on 1st October 2026.  The genderize.io name-to-gender data used in Fig.~\ref{F18} are licensed from genderize.io and are not redistributed.

\section{Author Contribution}
Simon J Porter: Conceptualization; Formal Analysis; Methodology; Visualisation; Writing - original draft; Writing - review \& editing. Daniel W Hook: Formal Analysis; Methodology; Visualisation; Writing - original draft; Writing - review \& editing.

\section{Conflict of interests}
The authors of this paper are both employees of Digital Science, the owner and operator of Dimensions.

\bibliography{Initials}
\clearpage
\end{document}